\documentclass{ws-ijbc}
\usepackage{graphicx}
\usepackage{epstopdf}
\usepackage{color}      
\begin{document}

\catchline{}{}{}{}{} 

\markboth{Skokos et al.}{Frequency map analysis of spatiotemporal chaos}

\title{FREQUENCY MAP ANALYSIS OF SPATIOTEMPORAL CHAOS IN THE NONLINEAR DISORDERED KLEIN-GORDON LATTICE}

\author{CHARALAMPOS SKOKOS}

\address{Nonlinear Dynamics and Chaos Group, Department of
    Mathematics and Applied Mathematics\\ University of Cape Town,
    Rondebosch, 7701, South Africa\\ haris.skokos@uct.ac.za,  haris.skokos@gmail.com}

\author{ENRICO GERLACH}
\address{Lohrmann Observatory,
  Technical University Dresden, D-01062, Dresden, Germany\\
enrico.gerlach@tu-dresden.de}

\author{SERGEJ FLACH}
\address{Center for Theoretical Physics of Complex Systems \\Institute for Basic Science, Daejeon 305-732, South Korea\\
sflach@ibs.re.kr}

\maketitle

\begin{history}
\received{(to be inserted by publisher)}
\end{history}

\begin{abstract}
We study the characteristics of chaos evolution of initially localized energy excitations in the one-dimensional nonlinear disordered Klein-Gordon lattice of anharmonic oscillators, by computing the time variation of the fundamental frequencies of the motion of each oscillator. We focus our attention on the dynamics of the so-called `weak' and `strong chaos' spreading regimes [Laptyeva et al., 2010, Europhys.~Lett., 91 30001], for which Anderson localization is destroyed, as the initially restricted excitation at the central region of the lattice propagates in time to more lattice sites. Based on the fact that large variations of the fundamental frequencies denote strong chaotic behavior, we show that in both regimes chaos is more intense at the central regions of the wave packet, where also the energy content is higher, while the oscillators at the wave packet's edges, through which the energy propagation happens, exhibit regular motion up until the time they gain enough energy to become part of the highly excited portion of the wave packet. Eventually, the percentage of chaotic oscillators remains practically constant, despite the fact that the number of excited sites grows as the wave packet spreads, but the portion of highly chaotic sites decreases in time. Thus, albeit the number of chaotic oscillators is constantly growing the strength of their chaotic behavior decreases, indicating that although chaos persists it is becoming weaker in time. We show that the extent of the zones of regular motion at the edges of the wave packet in the strong chaos regime is much smaller than in the weak chaos case. Furthermore, we find that in the strong chaos regime the chaotic component  of  the wave packet is not only more extended than in the weak chaos one, but in addition the fraction of strongly chaotic oscillators is much higher. Another important difference between the weak and strong chaos regimes is that in the latter case a significantly larger number of frequencies is excited, even from the first stages of the evolution. Moreover, our computations confirmed the shifting of fundamental frequencies outside the normal mode frequency band of the linear system in the case of the so-called `selftrapping' regime where a large part of the wave packet remains localized.
\end{abstract}



\section{Introduction}\label{sec:intro}

The nonlinear  disordered Klein-Gordon (DKG) lattice of coupled anharmonic oscillators has been extensively used in studies of the effect of nonlinearity on the energy propagation in disordered media, mainly in one (1D) \cite{FKS09,SKKF09,LBKSF10,SF10,F10,BLSKF11,BLGKSF11,SGF13,ABSD14,ASBF17,SMS18,SS21} but also in two (2D) \cite{LBF12,LIF14,MSS20} spatial dimensions. In these studies numerical evidences of the destruction of `Anderson localization' \cite{A58,KM93} (i.e.~the halt of energy spreading observed in linear, sufficiently disordered systems) were presented.

In particular, for the 1D DKG system, which has been studied in more detail than its 2D counterpart whose long-time numerical integration is a very demanding computational task, the existence of two energy spreading dynamical regimes was theoretically predicted and numerically verified: the so-called \emph{`weak'} and \emph{`strong chaos'} regimes, which are characterized by different dynamical behaviors. More specifically, the second moment $m_2$ of the normalized energy distribution increases in time $t$ as $m_2(t) \propto t^{a_m}$ with $a_m=1/3$ ($a_m=1/2$) for the weak (strong) chaos case, while the corresponding participation number $P$ grows as $P(t) \propto t^{a_p}$ with $a_p=1/6$ ($a_p=1/4$). In addition, the appearance of a  \emph{`selftrapping'} regime, where the bulk of the wave packet remains localized, was also theorized and observed \cite{KKFA08,FKS09,SKKF09,LBKSF10,F10,BLSKF11}. The generality of these results was substantiated by the fact that the same dynamical behaviors were observed also for the disordered discrete nonlinear Schr\"{o}dinger (DDNLS) equation \cite{M98,PS08,GS09,FKS09,SKKF09,LBKSF10,BLSKF11,BLGKSF11,SMS18,KYF20}, whose numerical integration is much more demanding from a computational point of view. Further substantiation was produced from high speed iterations of related nonlinear unitary maps using discrete time quantum walks which not only confirmed the generality of the spreading results, but which also allowed to push time horizons of subdiffusion to record $10^{12}$ \cite{VFF19}.

All these studies indicated the chaotic nature of energy spreading in nonlinear disordered media. The characteristics of this chaotic behavior were investigated in \cite{SGF13,SMS18} and \cite{MSS20} for, respectively, 1D and 2D disordered lattices, through  the time evolution of the most commonly used chaos indicator, the maximum Lyapunov exponent (MLE) $\Lambda_1$ \cite{BGGS80a,BGGS80b,S10,PP16} and the study of the properties of the so-called \emph{`deviation vector distribution'} (DVD) of the tangent space vector used for the computation of the MLE. There it was shown that the overall strength of chaos is decreasing as an initially localized energy excitation spreads to more lattice sites, but without any sign of a potentially crossover to regular behavior as was speculated in \cite{JKA10,A11}, at least up to the maximum evolution times reached in those studies. This behavior is reflected in the time decay of the finite time MLE (ftMLE) $\Lambda$, i.e.~$\Lambda(t) \propto t^{a_{\Lambda}}$, with $a_{\Lambda} \approx -0.25$ ($a_{\Lambda} \approx -0.3$) for the weak (strong) chaos regime of 1D systems \cite{SGF13,SMS18}, and with $a_{\Lambda} \approx -0.37$ ($a_{\Lambda} \approx -0.46$) for the weak (strong) chaos case of 2D systems \cite{MSS20}. It is also worth noting the existence of a dimension-independent scaling between the wave packet's spreading and chaoticity, quantified by the same behavior of the ratio $\Lambda(t)/m_2(t)$ in 1D and 2D systems \cite{MSS20}.

The scope of the current paper is to shed some additional light on the spatiotemporal characteristics of chaos in 1D disordered, nonlinear lattices. The MLE is an average measure of chaoticity providing information about the global behavior of a dynamical system. As such, the value of the MLE by itself is not enough to understand the spatiotemporal evolution of active chaotic regions in a multidimensional system \cite{TSL14}. Thus, in order to follow the chaotic or regular behavior of each individual lattice site in time we implement the so-called \emph{`frequency map analysis'} (FMA) technique \cite{Laskar1990, LFC1992, Laskar1993} and evaluate the  fundamental frequency  of an appropriate time series produced by the coordinates of each oscillator. Then, we use the variations of these frequencies to identify chaotic behavior. This approach allows us to visualize the evolution of chaos in the propagated energy distribution and to identify differences between the weak and strong chaos spreading regimes.

The paper is organized as follows. In Sect.~\ref{sec:model} we present the Hamiltonian function of the 1D DKG model along with the basic quantities we use to characterize the energy spreading and the system's chaoticity. The numerical results of our study are reported in Sect.~\ref{sec:num_res}, initially for some weak chaos cases and afterwards for the strong chaos regime, followed by a brief discussion of the selftrapping behavior. Finally in  Sect.~\ref{sec:summary} we summarize our findings and discuss their significance.

\section{Lattice model and numerical techniques}\label{sec:model}

The Hamiltonian function of the 1D DKG lattice system of $N$ anharmonic oscillators is
\begin{equation}
H = \sum_{i = 1}^{N} \left[ \frac{p_i ^2}{2} + \frac{\tilde{\epsilon}_l }{2} q_i ^2 +
\frac{q_i ^4}{4} + \frac{1}{2W}\left( q_{i+1} - q_{i} \right) ^2 \right],
\label{eq:H_DKG}
\end{equation}
with $q_i$, $p_i$ being respectively the generalized position and momentum of site $i$. The disorder parameters $\tilde{\epsilon}_l$ take uncorrelated random values in the interval $\left[ \frac{1}{2}, \frac{3}{2}\right]$ following a uniform probability distribution, and $W>0$ determines the disorder strength. In our study we consider fixed boundary conditions, i.e.~$q_0=p_0=q_{N+1}=p_{N+1}=0$. The Hamiltonian \eqref{eq:H_DKG} is an integral of motion and its value $H$ (usually referred to as the system's energy $E$) remains constant in time.

We use the ABA864 symplectic integrator (SI) \cite{BCFLMM13} for numerically solving the equations of motion
\begin{equation}
\dot{q_i} = \frac{d q_i}{d t}= \frac{\partial H}{\partial p_i}  , \,\,\,\,\,
\dot{p_i} = \frac{d p_i}{d t}= -\frac{\partial H}{\partial p_i} , \,\,\,\,\, i=1,2,\ldots, N,
\label{eq:H_eqmot}
\end{equation}
of Hamiltonian \eqref{eq:H_DKG}, which has been proved to be a very efficient technique for this task \cite{SS18,DMMS19}. In particular, for the implementation of the ABA864 SI we split Hamiltonian \eqref{eq:H_DKG} in two integrable parts, namely the system's kinetic and potential energy, and use an integration time step $\tau=0.5$, which, in general, keeps the absolute value of the relative energy error
\begin{equation}
E_r=\left|\frac{H(t)-H(0)}{H(0)} \right|
\label{eq:Energy_error}
\end{equation}
below  $10^{-5}$. In our simulations we typically excite one or few sites at the central region of the lattice and follow the evolution of this energy excitation in time, taking care that the lattice is large enough so that the energy does not reach its boundaries until the end of the integration.

In order to analyze the characteristics of the energy spreading we define the normalized energy distribution
\begin{equation}
\xi_i = \left[\frac{p_i^2}{2} + \frac{\tilde{\epsilon}_i}{2} q_i^2 + \frac{q_i^4}{4} +
\frac{1}{4W} \left( q_{i + 1} - q_{i} \right)^2\right] \Big/ E, \,\,\,\,\, i=1,2,\ldots,N,
\label{eq:energy_dist}
\end{equation}
and compute the distribution's second moment
\begin{equation}
m_2 = \sum_{i = 1}^{N} (i - \bar{i})^2 \xi_i,
\label{eq:m2}
\end{equation}
which measures the extent of the distribution, and the participation number
\begin{equation}
P = \left[ \sum_{i = 1}^{N} \xi_i^2 \right]^{-1},
\label{eq:P}
\end{equation}
which estimates the number of highly excited sites. We note that in \eqref{eq:m2} $\bar{i}=\sum_{i = 1}^{N} i \xi_i$ is the distribution's mean position.

The system's chaoticity is quantified by the MLE $\Lambda_1$, which is estimated as the limit for $t\rightarrow \infty$ of the ftMLE
\begin{equation}
\Lambda (t) = \frac{1}{t}\ln \frac{\| \vec{w}(t)
\|}{\| \vec{w}(0) \|},
\label{eq:ftMLE}
\end{equation}
i.e.~$\Lambda_1 = \lim_{t\to\infty} \Lambda(t)$ \cite{BGGS80a,BGGS80b,S10,PP16}. In \eqref{eq:ftMLE}
$\vec{w}(0)$ and $\vec{w}(t)$ are respectively phase space deviation vectors from the studied orbit at $t = 0$
and $t > 0$,  while $ \| \cdot \|$ denotes the usual Euclidian norm of a vector. The evolution of the deviation vector is governed by the so-called \emph{`variational equations'} (see e.g.~\cite{S10} for more details), whose integration is done according to the \emph{`tangent map method'} \cite{SG10,GS11,GES12} using the ABA864 SI.

Here we are interested in the spatiotemporal evolution of chaos in the DKG system \eqref{eq:H_DKG}. Thus, since the MLE provides information about the collective and overall chaotic behavior of the system, we implement the FMA method for identifying the time evolution of the chaotic behavior of individual lattice sites. Neglecting the non-linear term $q^4/4$ in the Hamiltonian~\eqref{eq:H_DKG}, the system is integrable and the motion of each site is fully determined by a certain linear combination of the eigenfrequencies of this system. Under the influence of the non-linearity these frequencies will become time-dependent and can be determined by using a refined Fourier technique as for example the NAFF (\textit{Numerical Analysis of Fundamental Frequencies}) algorithm. For a detailed description of NAFF and its application for FMA in various research fields see for example \cite{Laskar1990, Laskar1993, RL2000, Y2014} and references therein.

This algorithm uses as input the numerically integrated trajectory $q_i(t),\, p_i(t)$ of oscillator $i$ over a time span $T$ and identifies the fundamental frequenciesin the quasiperiodic approximation of this motion. In this work we compute the relative change of the largest of these fundamental frequencies $f_{1i}$ and $f_{2i}$ between two successive time windows of length $T=6\cdot 10^5$ time units.\footnote{During initial test runs we found that $T=6\cdot 10^5$ time units provides reliable and accurate estimates for $f$.}  In the case of regular, quasiperiodic motion the two frequencies should theoretically be equal (in practice they slightly differ due to their numerical estimation), while for chaotic motion the two computed frequencies are in general different, as there is no reason for them to remain constant. Thus, the relative change of these two frequencies, quantified by
\begin{equation}
D_i=\left| \frac{f_{2i}-f_{1i}}{f_{1i}} \right|,
\label{eq:D}
\end{equation}
can be used to identify the chaotic or regular nature of motion. More specifically, small $D_i$ values denote the practical constancy of the fundamental frequency and consequently indicate regular motion, while large $D_i$ values signify chaotic behavior characterized by strong variations in the computed frequency values.

\section{Numerical results}\label{sec:num_res}

Although in our analysis we considered several individual cases belonging to the weak and strong chaos regimes, as well as the selftrapping regime (more information on the definition of these dynamical behaviors can be found in \cite{LBKSF10,F10}), we will only report our findings for some representative cases, which have already being studied in the literature. In all cases we consider a disorder realization, i.e.~a set of $\tilde{\epsilon}_l$ values in \eqref{eq:H_DKG}, and  excite $L$ central sites of a lattice of $N$ oscillators by setting their positions at zero and adjusting their momenta so that each one of them has the same energy density $e=E/L$. For single site excitations ($L=1$) always a positive momentum is used, while for
block excitations ($L>1$) the sign of the initial momentum of each excited site is chosen randomly. In particular, we  consider the following cases:
\begin{enumerate}
  \item Weak chaos regime
    \begin{enumerate}
      \item Case WCI (single site excitation): $N=999$, $W=4$, $L=1$,  $e=E=0.4$ (case W2$_K$ of \cite{SMS18}).
      \item Case WCII (block excitation): $N=999$,  $W=4$, $L=21$,  $e=0.01$,  $E=0.21$ (case W3$_K$ of \cite{SMS18}).
    \end{enumerate}
  \item Strong chaos regime
    \begin{enumerate}
      \item Case SCI (block excitation): $N=3499$,  $W=4$, $L=21$,  $e=0.2$,  $E=4.2$ (case studied in \cite{BLSKF11}).
      \item Case SCII (block excitation): $N=2499$, $W=3$, $L=37$,  $e=0.1$,  $E=3.7$ (case S2$_K$ of \cite{SMS18}).
    \end{enumerate}
  \item Selftrapping regime
    \begin{enumerate}
      \item Case STI (single site excitation): $N=399$, $W=4$, $L=1$,  $e=E=1.5$ (case studied in \cite{SKKF09}).
    \end{enumerate}
\end{enumerate}

\subsection{Weak chaos dynamical regime}\label{sec:num_weak}

\begin{figure}
\centering
\includegraphics[width=\textwidth]{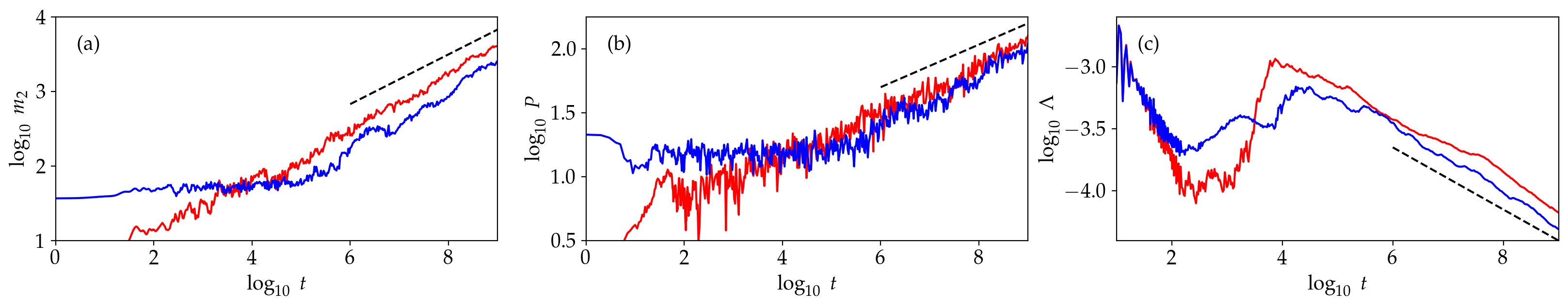}
\caption{Time evolution of (a) the second moment $m_2(t)$ \eqref{eq:m2}, (b) the participation number  $P(t)$ \eqref{eq:P}, and (c) the ftMLE $\Lambda(t)$ \eqref{eq:ftMLE} for the weak chaos cases WCI (red curves) and WCII (blue curves) in log-log scale. The straight dashed lines indicate slopes (a) $1/3$, (b) $1/6$ and (c) $-0.25$. }
\label{fig:wc1_m2pL}
\end{figure}

In Fig.~\ref{fig:wc1_m2pL} we see that the evolution of the second moment $m_2(t)$ [Fig.~\ref{fig:wc1_m2pL}(a)], the participation number  $P(t)$ [Fig.~\ref{fig:wc1_m2pL}(b)] and the ftMLE $\Lambda(t)$ [Fig.~\ref{fig:wc1_m2pL}(c)] exhibits the expected behavior for both the WCI (red curves) and the WCII case (blue curves). More specifically, in accordance to \cite{FKS09,SKKF09,LBKSF10,F10,BLSKF11} the asymptotic behavior of the wave packet's dynamics is characterized by $m_2(t) \propto t^{1/3}$ and $P(t) \propto t^{1/6}$, while the ftMLE shows a decrease $\Lambda(t) \propto t^{-0.25}$ as was reported  in \cite{SGF13,SMS18}.

\begin{figure}
  \centering
  \includegraphics[width=\textwidth]{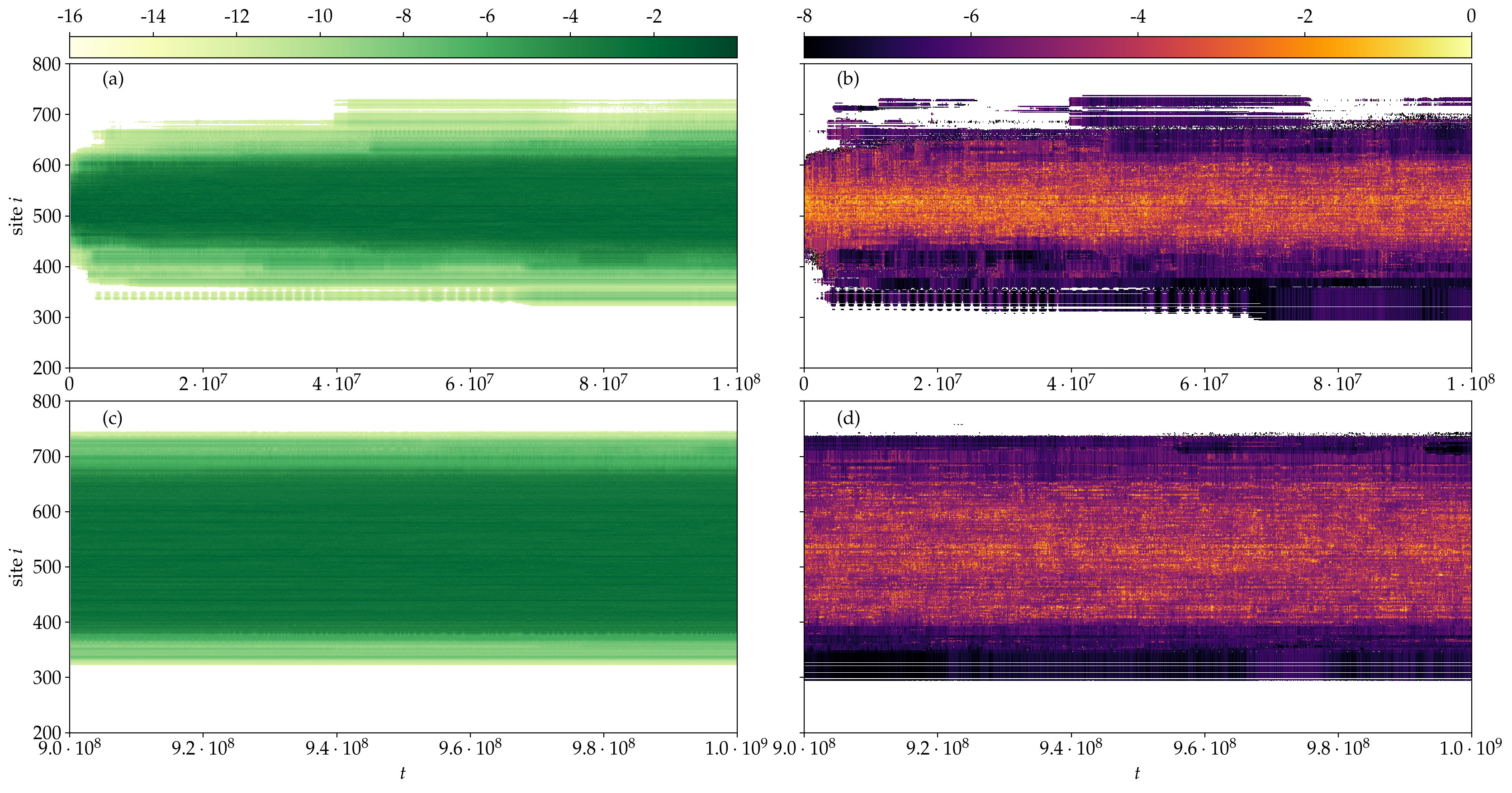}
\caption{The time evolution of [(a) and (c)] the normalized energy distribution $\xi_i$ \eqref{eq:energy_dist} and [(b) and (d)] the quantity $D_i$ \eqref{eq:D} for the WCI case in the time intervals [(a) and (b)] $0 \leq t \leq 10^8$ and [(c) and (d)] $9 \cdot 10^8 \leq t \leq 10^9$. In each panel the horizontal axis is time $t$ in linear scale, while the vertical axis shows the  site number $i$. The color scales at the top of the upper row of panels are used for coloring lattice sites according to the [(a) and (c)] $\log_{10} \xi_i$ and [(b) and (d)] $\log_{10} D_i$ values.}
\label{fig:wc1_energy_D}
\end{figure}

Let us first study in detail the WCI weak chaos case where initially 1 site is excited at the middle of a lattice with $N=999$ oscillators. As time increases the system's constant total energy $E$ \eqref{eq:H_DKG} is spread to more sites as can be seen from Figs.~\ref{fig:wc1_energy_D}(a) and (c) where  we depict the energy evolution in two time windows of length $10^8$ time units: immediately after the initial excitation [Fig.~\ref{fig:wc1_energy_D}(a)] and after $9 \cdot 10^8$ time units [Fig.~\ref{fig:wc1_energy_D}(c)]. In order to visualize the change in time of the chaoticity of each lattice site we also perform the FMA of the system and compute the evolution of the quantity $D_i$ \eqref{eq:D}, which captures the variation of the sites' fundamental frequencies in two successive time windows of length $6\cdot 10^5$  attributing the computed $D_i$ value of each site at the beginning of these two time intervals.  The outcome of this process is presented in Figs.~\ref{fig:wc1_energy_D}(b) and (d), respectively for the time intervals of Figs.~\ref{fig:wc1_energy_D}(a) and (c). There each site is colored according to its $\log_{10} D_i$ value. Chaotic behavior corresponding to large frequency changes and high $\log_{10} D_i$ values is indicated by red and yellow colors, while regions of low $\log_{10} D_i$ values denoting small frequency changes attributed to regular, non-chaotic behavior are colored in purple and black. From the results of   Figs.~\ref{fig:wc1_energy_D}(b) and (d) we clearly see that chaos appears at the central regions of the wave packet, where also the energy concentration is higher as  Figs.~\ref{fig:wc1_energy_D}(a) and (c) indicate. On the other hand, the edges of the wave packet, through which energy spreading takes place, behave more regularly having also significantly smaller energies than the central regions.

It is worth mentioning here a practical issue. In order to compute the change in the fundamental frequencies using the NAFF algorithm for two successive time windows of length $6\cdot 10^5$  time units, we have to ensure that the time series we analyze is reliable and accurate. For this reason we perform the computation of $D_i$ only if the absolute values for each the two successive frequencies $f_{1i}$  and $f_{2i}$ are above the adopted threshold of $10^{-16}$. If this condition is not satisfied then we do not regard the computed frequency as accurate and we dismiss it. For the computation of $D_i$ \eqref{eq:D} we require the obtainment of the fundamental frequency in two successive time windows. If, according to the above-mentioned criterion, this is not possible for either of these time intervals we do not compute $D_i$. Consequently, there are several points in Figs.~\ref{fig:wc1_energy_D}(b) and (d), mainly at the edges of the wave packet, where the values of $|f_{1i}|$ and/or $|f_{2i}|$   are very small, for which we do not compute a $D_i$ value, although these sites have (very small) energies, as can be seen in Figs.~\ref{fig:wc1_energy_D}(a) and (c).

Although the color maps of $D_i$ values in Figs.~\ref{fig:wc1_energy_D}(b) and (d) accurately capture subtle changes in the chaotic behavior of lattice sites, we  also follow below a different approach in order to clearly visualize the spatiotemporal evolution of chaos in the system. In order to do so, let us first note that the frequencies of the normal modes of the linear DKG model, i.e.~the system obtained by neglecting the nonlinear terms $q_i^4/4$ in Hamiltonian \eqref{eq:H_DKG}, belong to a well-defined interval. More specifically, the normal mode eigenvalues $\omega_{\nu}^2 \in \left[ \frac{1}{2}, \frac{3}{4} + \frac{4}{W} \right]$, $\nu=1, 2, \ldots, N$ (see e.g.~\cite{SPMS20} and references therein), so that
\begin{equation}
f_{\nu} =\frac{\omega_{\nu}}{2 \pi} \in \left[ \frac{1}{2 \sqrt{2} \pi}, \frac{\sqrt{3W+8}}{2 \pi \sqrt{2W}} \right].
\label{eq:fn_boundaries}
\end{equation}
The presence of nonlinearity induces a  frequency shift to higher frequency values \cite{SF10}. This shift  increases with the growth of the system's energy $E$, which actually plays the role of nonlinearity's strength. Thus, the fundamental frequencies of all lattice oscillators are expected to belong to a frequency band similar to \eqref{eq:fn_boundaries}, having somehow raised lower and upper limits. This is actually true as our numerical results reported below clearly verify.

In our study we consider set-ups for only two values of $W$, namely $W=3$ (case SCII) and $W=4$ (all other cases). For each $W$ value we consider a range of possible frequencies, which is slightly larger than the one reported in \eqref{eq:fn_boundaries} in order to be certain that it will contain the actually computed frequencies of the nonlinear system \eqref{eq:H_DKG}. In particular, for $W=3$ we consider the frequency  band $\left[ 0.1, 0.28 \right]$ and for $W=4$ the interval    $\left[ 0.1, 0.26 \right]$. These ranges are then divided into $n=500$ bins of equal length, setting the width of each frequency bin to be of the order of $10^{-4}$.  During the wave packet's evolution the fundamental frequencies are computed and for each site the bin which contains the found frequency is registered. This is called \textit{one visit} to that bin. In the course of the integration, the number of visits to the bin that received the most visits up to this point for each site is divided by the number of total registered bin visits (i.e.~total number of frequency computations) for this site. This gives the \textit{relative number of visits} of the most visited frequency bin for each site and is used to measure what we call \emph{`frequency locking'} of site $i$ ($FL_i$). This number is always $0\le FL_i \le 1$. The value $FL_i=1$ for a particular site $i$ indicates that during the evolution this site had always the same frequency (actually this means that although the frequency might change a bit it remained in the same bin, i.e.~it does not change much as the bins' width is $\sim 10^{-4}$), signifying effectively regular behavior. On the other hand, small $FL_i$ values denote many changes of the fundamental frequency as the placement of the computed values in one frequency bin is low, and consequently indicate chaotic behavior, which becomes stronger as $FL_i$ decreases.

The computation of $FL_i$ is done based on data from a time window of $1.2\cdot 10^6$ time units and repeated each $1.2\cdot 10^5$ time units. To avoid saturation effects the computation is re-initialized after a fixed time interval, which is set to be $t_{FL}=2.4\cdot 10^7$ in order to better follow the time evolution of this quantity in different evolution stages of the wave packet. By doing that we would for example be able to identify potential regularly behaving epochs in the dynamics of a lattice site, i.e.~time intervals for which its frequency remains practically constant, although from time to time the value of the practically constant frequency might change. 

\begin{figure}
\centering
\includegraphics[width=\textwidth]{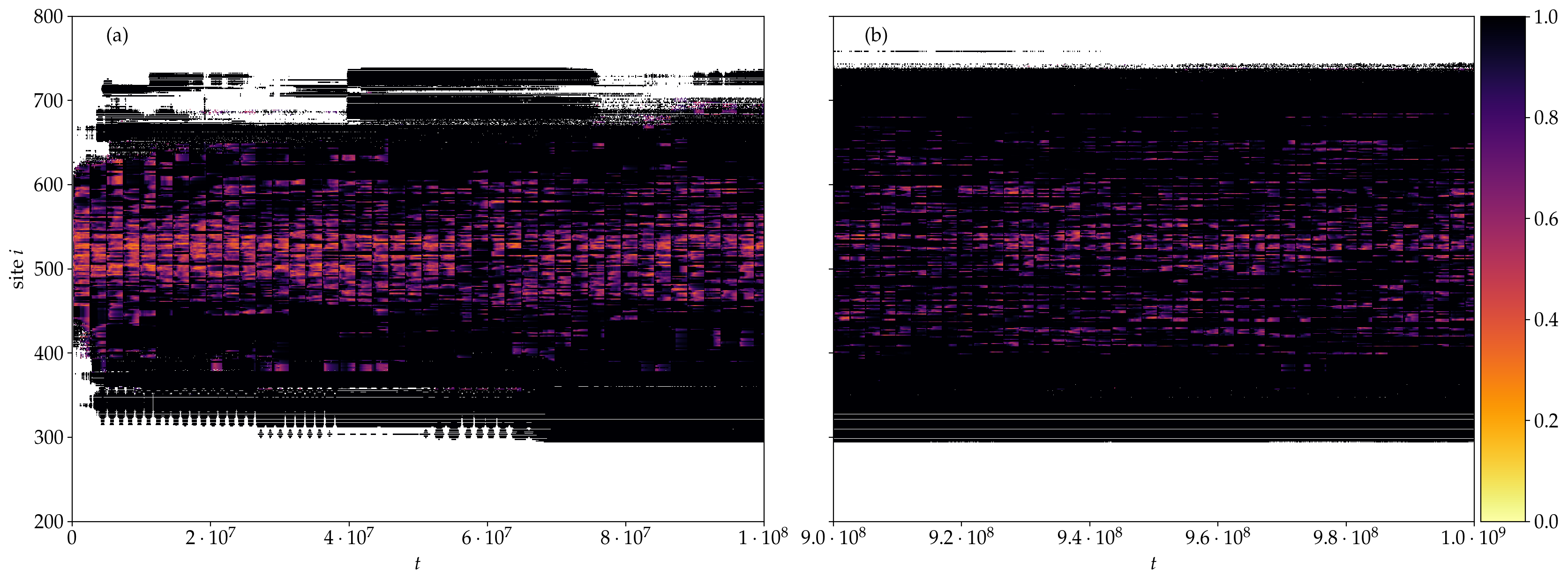}
\caption{Frequency locking $FL_i$ values at each lattice site $i$ for (a) $0 \leq t \leq 10^8$ and (b) $9 \cdot 10^8 \leq t \leq 10^9$ (WCI case). The color scales on the right of the panels are used to color the points for which the fundamental frequencies were obtained, according to their $FL_i$ values.}
\label{fig:wc1_FL}
\end{figure}

The results of the computation of $FL_i$ for the WCI case are shown in Figs.~\ref{fig:wc1_FL}(a) and (b) respectively for the time intervals $0 \leq t \leq 10^8$ and $9 \cdot 10^8 \leq t \leq 10^9$. Points for which the fundamental frequencies of the lattice sites were computed are colored according to their $FL_i$ values, with red and yellow colors denoting small $FL_i$ values indicating strong chaotic behavior. On the other hand, regions with very high $FL_i$ values close to $1$ colored in black correspond to regular motion. The re-initialization of the $FL_i$'s computation every $t_{FL}=2.4\cdot 10^7$ time units results in the creation of the equidistant vertical `lines' in both panels of   Fig.~\ref{fig:wc1_FL}.

In accordance to the results presented in Figs.~\ref{fig:wc1_energy_D}(b) and (d), Fig.~\ref{fig:wc1_FL} shows that the edges of the wave packet behave regularly, while chaos appears at the central region of the lattice's excited part. The fact that in the $FL_i$ plots of Fig.~\ref{fig:wc1_FL} we do not restrict ourselves in computing the exact frequencies, but we allow some frequency changes of the order of the width of the bin, i.e. $\sim 10^{-4}$, results in the creation of pictures depicting the system's dynamics more clearly than the $D_i$ color plots [Figs.~\ref{fig:wc1_energy_D}(b) and (d)], emphasizing especially the more active chaotic regions of the lattice, which exhibit large changes in their frequencies. The $D_i$ plots are more detailed and of course more accurate, but the many details they provide make the picture somewhat more obscure, hiding the main dynamical features of the system. For example, in Fig.~\ref{fig:wc1_FL}(b) the footprint of a horizontal `stripe' of regular motion around $i \approx 480$, lasting for almost the whole duration of the presented time interval, is easily identified as points there are mainly colored in black. This region is also present in Fig.~\ref{fig:wc1_energy_D}(d) where points are mostly colored in purple/dark red, indicating values of $\log_{10}D_i \approx 10^{-5}$, but is not as easily visible as in  Fig.~\ref{fig:wc1_FL}(b).

\begin{figure}
\centering
\includegraphics[width=\textwidth]{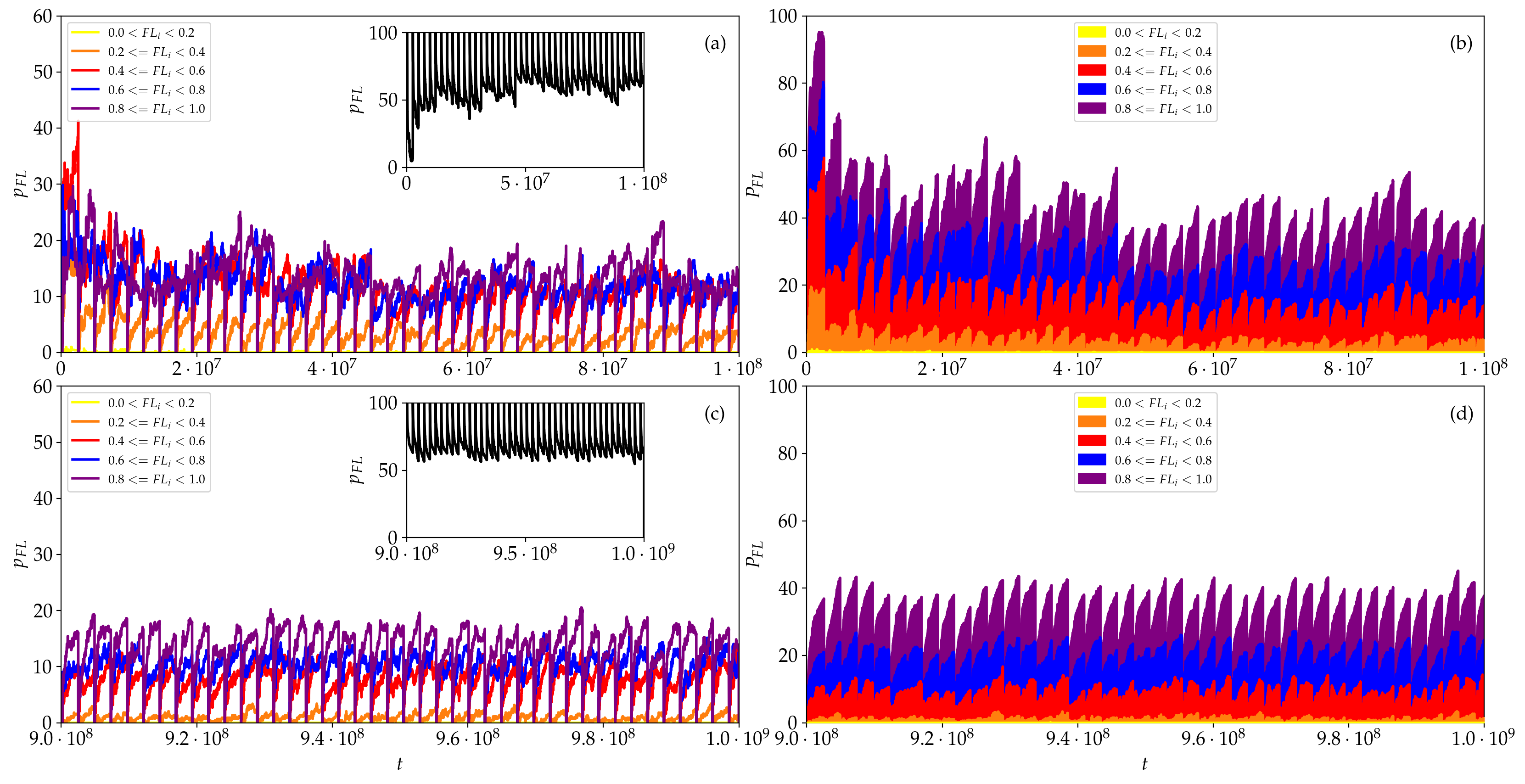}
\caption{Time evolution of [(a) and (c)] the percentages $p_{FL}$ of sites exhibiting $FL_i$ values in particular intervals,  and [(b) and (d)] the accumulated percentages $P_{FL}$ of sites with values in a particular $FL_i$ range, for [(a) and (b)] $0 \leq t \leq 10^8$  and [(c) and (d)] $9 \cdot 10^8 \leq t \leq 10^9$ (WCI case). The considered intervals are: $0< FL_i <0.2$ (yellow curves), $0.2 \leq FL_i <0.4$ (orange curves), $0.4 \leq FL_i <0.6$ (red curves), $0.6 \leq FL_i <0.8$ (blue curves), $0.8 \leq FL_i < 1$ (purple curves). Insets in panels (a) and (c): The time evolution of the percentage of sites with $FL_i=1$, i.e.~points plotted in black in Fig.~\ref{fig:wc1_FL}, which correspond to regions of regular motion. All presented  results are obtained by the analysis of the outcomes presented in Fig.~\ref{fig:wc1_FL}. The re-initialization of the $FL_i$'s computation every $t_{FL}=2.4\cdot 10^7$  time units leads to the vertical `stripe'-looking feature appearing in all panels. 
}
\label{fig:wc1_FL_percent}
\end{figure}

The comparison of Figs.~\ref{fig:wc1_FL}(a) and (b) shows that as time grows the portion of the wave packet exhibiting  chaotic behavior (i.e.~points not colored in black) decreases. This is mainly due to the fact that the regularly behaving sites at the black colored edges of the wave packet grow. In order to quantify this observation we plot in Figs.~\ref{fig:wc1_FL_percent}(a) and (c) the evolution of the percentage $p_{FL}$ of sites, respectively shown in Figs.~\ref{fig:wc1_FL}(a) and (b),  having $FL_i$ values in various intervals. In particular, we consider the intervals $0< FL_i <0.2$ (yellow curves), $0.2 \leq FL_i <0.4$ (orange curves), $0.4 \leq FL_i <0.6$ (red curves), $0.6 \leq FL_i <0.8$ (blue curves), $0.8 \leq FL_i < 1$ (purple curves), while in the insets of Figs.~\ref{fig:wc1_FL_percent}(a) and (c) we plot the evolution of $p_{FL}$ for sites having $FL_i=1$, i.e.~sites whose fundamental frequency does not change much as it always remains in one bin indicating regular motion. The presented results are obtained by the analysis of the data shown in Fig.~\ref{fig:wc1_FL}, so we see again here a vertical `stripe' formation of length $t_{FL}=2.4\cdot 10^7$  related to the re-initialization of the $FL_i$ computations. In Figs.~\ref{fig:wc1_FL_percent}(b) and (d) we depict the accumulated percentages $P_{FL}$ obtained by adding up respectively the results of Figs.~\ref{fig:wc1_FL_percent}(a) and (c), with data belonging to the various ranges of $FL_i$ values colored according to the color code used in Figs.~\ref{fig:wc1_FL_percent}(a) and (c) [these color ranges are also given in the legends of Figs.~\ref{fig:wc1_FL_percent}(b) and (d)].

From the results of Figs.~\ref{fig:wc1_FL_percent}(a) and (b) we see that at the initial stages of the evolution the percentage of chaotic sites (i.e.~the ones having $FL_i \neq 1$) decreases, and obviously the number of sites showing consistent regular behavior ($FL_i = 1$) increases [see inset of Fig.~\ref{fig:wc1_FL_percent}(a)]. It is worth noting that the percentage of strongly chaotic regions, i.e.~the ones having $FL_i < 0.4$, remains relative low, e.g.~$p_{FL} \approx 3.4\%$  for sites with $0.2 \leq FL_i <0.4$ [orange curve in Fig.~\ref{fig:wc1_FL_percent}(a)], while the $p_{FL}$ values for $FL_i <0.2$ are too small ($\approx 0.02\%$)  so that the corresponding yellow curve is practically not visible in Fig.~\ref{fig:wc1_FL_percent}(a). Furthermore, the fraction of sites of moderate  ($0.4 \leq FL_i <0.6$ and  $0.6 \leq FL_i <0.8$ - red  and blue curves respectively) and of weak chaotic behavior ($0.8 \leq FL_i < 1$  - purple curve) show signs of saturation to values $p_{FL} \approx 10 \%$  for each one of these three $FL_i$ ranges. This saturation to some asymptotic $p_{FL}$ and $P_{FL}$ values becomes evident in Figs.~\ref{fig:wc1_FL_percent}(c) and (d). More specifically, we see that the red, blue and purple curves in Fig.~\ref{fig:wc1_FL_percent}(c) oscillate around values $p_{FL} \approx 10 \%$  with red (purple) curves being slightly below (above) this value. Furthermore, the percentages of well-established regular behavior remains practically constant around $p_{FL} \approx 65 \% $  [inset of Fig.~\ref{fig:wc1_FL_percent}(c)], while the strong chaotic behavior ($FL_i < 0.4$), corresponding to orange and yellow curves [which are practically not visible in Fig.~\ref{fig:wc1_FL_percent}(c)], obtains very small values $p_{FL} \lesssim 1 \% $. 

From the results of Figs.~\ref{fig:wc1_energy_D}--\ref{fig:wc1_FL_percent}  we see that chaotic behavior indicated by high $D_i$ values [points colored in red and yellow Figs.~\ref{fig:wc1_energy_D}(b), (d)] and by low $FL_i$ values [points not colored black in Fig.~\ref{fig:wc1_FL} and data presented in the main panels of Fig.~\ref{fig:wc1_FL_percent}], are mainly concentrated at the central regions of the wave packet. These results suggest that chaoticity is not extended, as the large portion of sites at the edges of the wave packet, whose frequency does not practically change, exhibit regular behavior. Thus, in general, points which are slightly excited at the wave packet's boundaries  are initially not chaotic (having relative small $D_i$ and large $FL_i$ values) until they become part of  the highly excited component of the wave packet, when more energy reaches them.

\begin{figure}
\centering
\includegraphics[width=\textwidth]{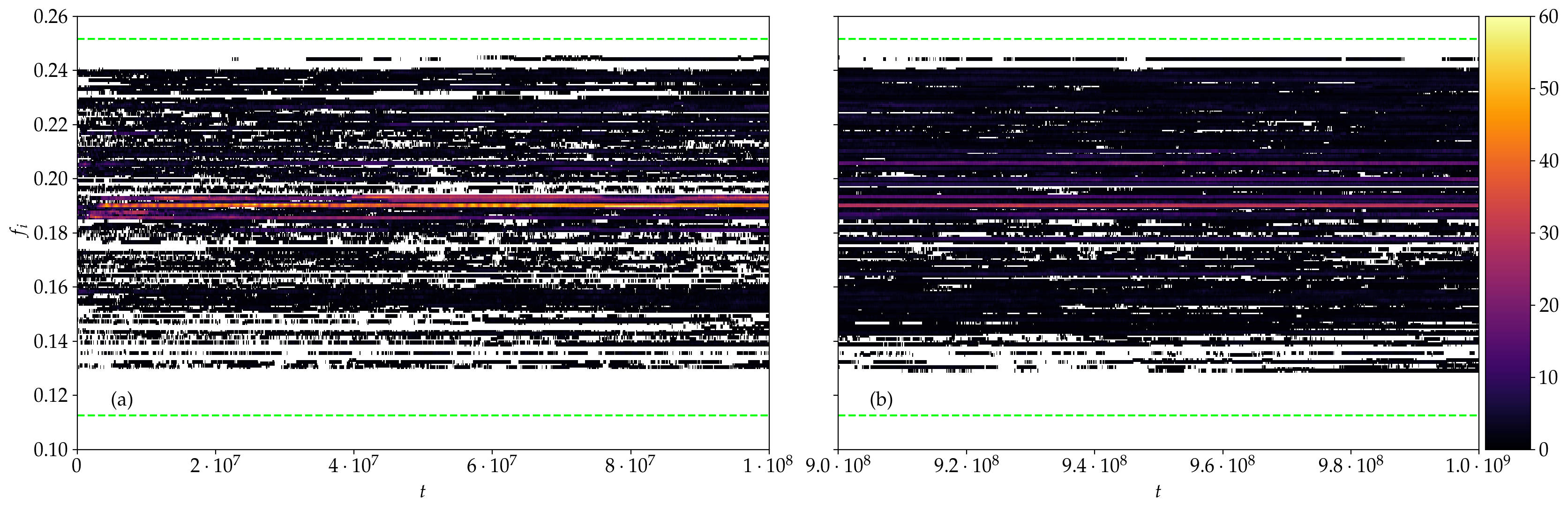}
\caption{Time evolution of the computed site frequencies $f_i$ in the WCI case for the time intervals (a) $0 \leq t \leq 10^8$ and (b) $9 \cdot 10^8 \leq t \leq 10^9$. The presented frequency range $[0.1, \, 0.26]$ has been divided in $n=500$ bins of equal length and each bin is colored according to the number of sites having frequencies in its value range at each time instant [which quantifies the bin's `frequency strength' ($FS$)] according to the color scales at the right side of the panels. White colored areas correspond to regions where no frequencies were found. The horizontal thick dashed green lines indicate the frequency band borders \eqref{eq:fn_boundaries} of the linear DKG system. 
}
\label{fig:wc1_bin}
\end{figure}

Furthermore, during the above-mentioned $FL_i$ computations additional information is gathered. More specifically, for each frequency bin the number of sites having their fundamental frequency in the bin is registered. The intention here  is to identify frequencies (or more accurately frequency bins, i.e.~small frequency ranges) which dominate the dynamics of the system by having many sites oscillating with these particular frequencies. In this way we quantify the importance or the influence of the particular frequency range on the dynamical evolution of the lattice.  The outcome of these computations are shown in Fig.~\ref{fig:wc1_bin} for the  two time intervals  $0 \leq t \leq 10^8$ [Fig.~\ref{fig:wc1_bin}(a)] and $9 \cdot 10^8 \leq t \leq 10^9$ [Fig.~\ref{fig:wc1_bin}(b)] we have considered so far. If at a given time the computed fundamental frequencies of some sites are located inside a particular frequency bin then a point marks this occurrence in Fig.~\ref{fig:wc1_bin}. This point is colored according to the number of sites belonging in the frequency bin at each time instant (we will refer to this measure as the \emph{`frequency strength'}, $FS$, of the particular frequency bin) following the color code depicted at the right side of Fig.~\ref{fig:wc1_bin}. Thus, red and yellow colors indicate many sites in a particular frequency
range, while the lack of computed frequencies correspond to white regions in  Fig.~\ref{fig:wc1_bin}.

From the results of Fig.~\ref{fig:wc1_bin} we see that as time grows more frequencies are excited since  the extent of the white regions in Fig.~\ref{fig:wc1_bin}(b) is  smaller that the one observed in  Fig.~\ref{fig:wc1_bin}(a). Furthermore, in both time windows only a few frequency bins are highly populated (the ones colored in red and yellow). The corresponding frequencies are mainly located near the middle of the frequency band of Fig.~\ref{fig:wc1_bin} ($f_i \approx 0.19$),  although some frequencies with higher $FS$ values are also present in the interval $f_i= 0.2$--$0.22$ [especially in  Fig.~\ref{fig:wc1_bin}(b)].

\begin{figure}
\centering
\includegraphics[width=\textwidth]{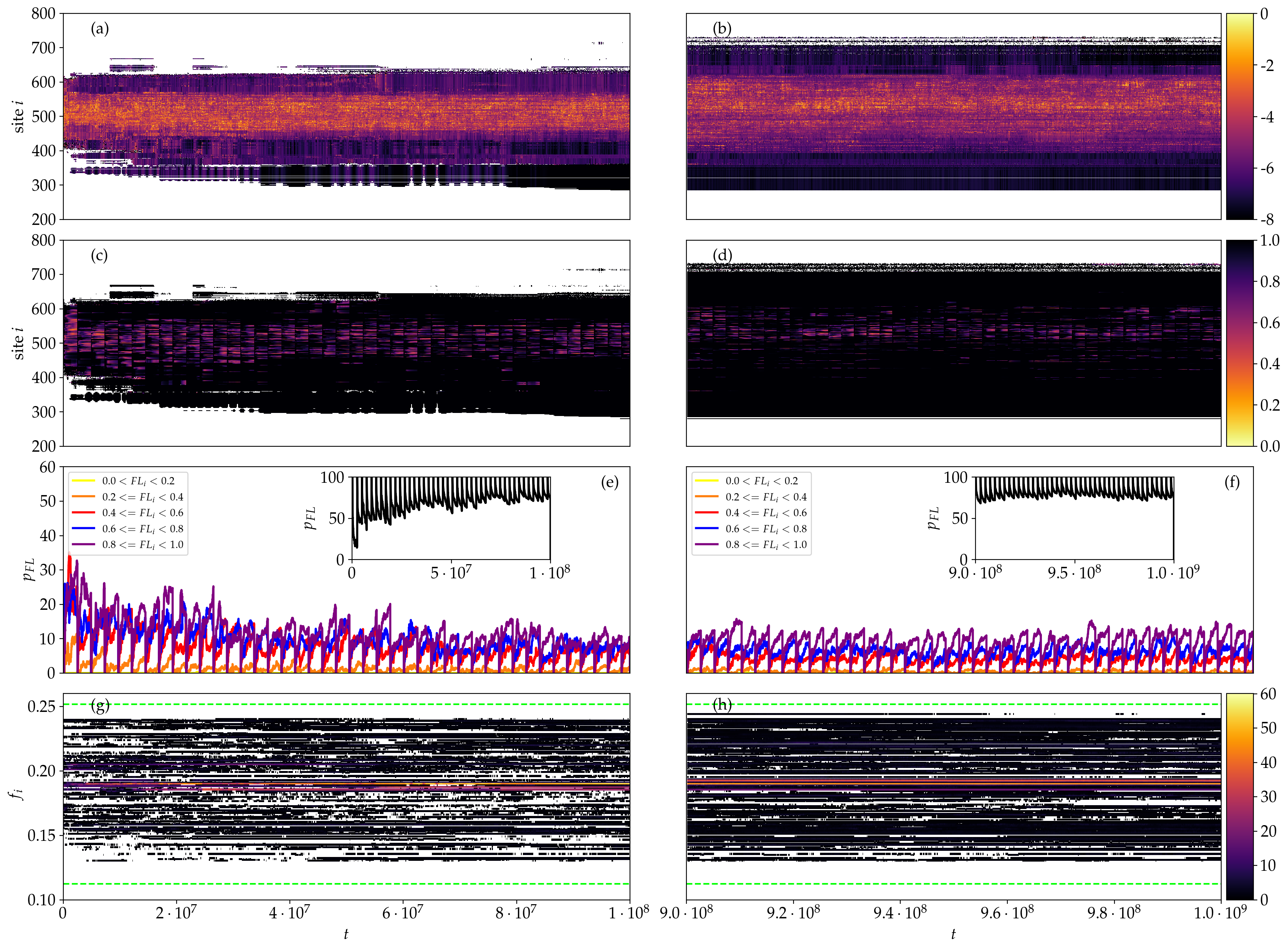}
\caption{Weak chaos case WCII. Time evolution of various quantities computed during the wave packet propagation in the time windows [(a), (c), (e), (g)] $0 \leq t \leq 10^8$  and  [(b), (d), (f), (h)] $9 \cdot 10^8 \leq t \leq 10^9$: [(a) and (b)] the quantity $D_i$ \eqref{eq:D} [plots similar to Figs.~\ref{fig:wc1_energy_D}(b) and (d)], [(c) and (d)] frequency locking $FL_i$ values (plots similar to Fig.~\ref{fig:wc1_FL}), [(e) and (f)] percentages $p_{FL}$ [plots similar to Figs.~\ref{fig:wc1_FL_percent}(a) and (c)], and [(g) and (h)] computed site frequencies $f_i$ (plots similar to Fig.~\ref{fig:wc1_bin}).
}
\label{fig:wc2}
\end{figure}

The results presented in Figs.~\ref{fig:wc1_energy_D}--\ref{fig:wc1_bin} for the weak chaos case WCI are quite general and characteristic of the weak chaos spreading regime, as similar behaviors have been observed for all other studied weak chaos cases. As a testimony to that we present in Fig.~\ref{fig:wc2} results obtained for the WCII case in which we initially excite $L=21$ neighboring lattice sites, in contrast to the single site excitation ($L=1$) of case WCI. For the WCII case we see again that as the wave packet spreads the region of intense chaoticity [i.e.~points colored in red and yellow in the $D_i$ plots in Figs.~\ref{fig:wc2}(a) and (b)] remains confined, well inside the extent of the wave packet, at the borders of which we encounter extended regions of regular behavior [black colored areas in Figs.~\ref{fig:wc2}(c) and (d)]. Again the percentage $p_{FL}$ of sites exhibiting complete frequency locking ($FL_i=1$) eventually remains practically  constant attaining rather high values, $p_{FL} \approx 80\%$, as can be seen in the insets of  Figs.~\ref{fig:wc2}(e) and (f), which are somewhat larger than the ones observed in the WCI case [$p_{FL} \approx 60\%$ - see insets of Figs.~\ref{fig:wc1_FL_percent}(a) and (c)]. At the same time, as in the WCI case, the fraction of the highly chaotic sites characterized by $FL_i <0.4$ is very small,  $p_{FL} \approx 0.5\%, $ with the remaining sites behaving chaotically, i.e.~having $FL_i$ values in the range $[0.4, \, 1)$ [Figs.~\ref{fig:wc2}(e) and (f)]. Figs.~\ref{fig:wc2}(g) and (h) clearly show that, as in the WCI case (Fig.~\ref{fig:wc1_bin}), initially  a few frequencies are excited, but later on, as the energy spreads to more sites,  the number of excited frequencies increases. Again a rather small number of frequencies, mainly located near the middle of the frequency band [points colored in red and yellow in Figs.~\ref{fig:wc2}(g) and (h)] strongly influence the dynamics of the lattice as they are highly populated for  most of the integration time.

\subsection{Strong chaos dynamical regime}\label{sec:num_strong}

\begin{figure}
\centering
\includegraphics[width=\textwidth]{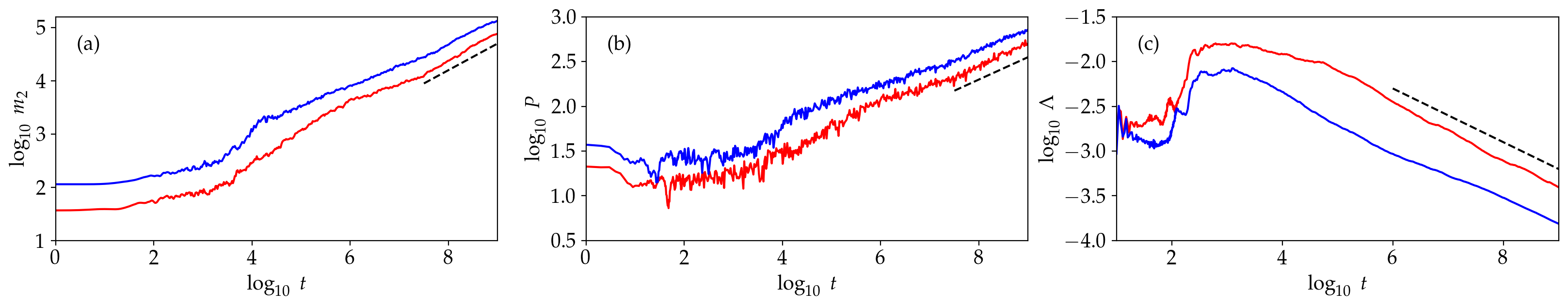}
\caption{Similar to Fig.~\ref{fig:wc1_m2pL} but for the strong chaos cases SCI (red curves) and SCII (blue curves): time evolution of (a) the second moment $m_2(t)$ \eqref{eq:m2}, (b) the participation number  $P(t)$ \eqref{eq:P}, and (c) the ftMLE $\Lambda(t)$ \eqref{eq:ftMLE} in log-log scale. The straight dashed lines indicate slopes (a) $1/2$, (b) $1/4$ and (c) $-0.3$.  }
\label{fig:sc1_m2pL}
\end{figure}

Let us now focus on the strong chaos spreading dynamical regime, which is characterized by a potentially long-lasting faster expansion of the wave packet, with respect to what is observed in the weak chaos case, eventually followed by a crossover to a slower subdiffusive weak chaos spreading \cite{LBKSF10,F10,BLSKF11,SMS18}. In Fig.~\ref{fig:sc1_m2pL} we present results for the two representative strong chaos cases we consider in our study, namely cases SCI (red curves) and SCII (blue curves).  All quantities are eventually behaving according to what is expected in the strong chaos regime. More specifically the second moment \eqref{eq:m2} and the participation number  $P(t)$ \eqref{eq:P} follow respectively power law growths $m_2(t) \propto t^{1/2}$ and $P(t)\propto t^{1/4}$ \cite{LBKSF10,F10,BLSKF11}, while the decrease of the  ftMLE  \eqref{eq:ftMLE} is well described by $\Lambda(t) \propto t^{-0.3}$ \cite{SMS18}.

\begin{figure}
\centering
\includegraphics[width=\textwidth]{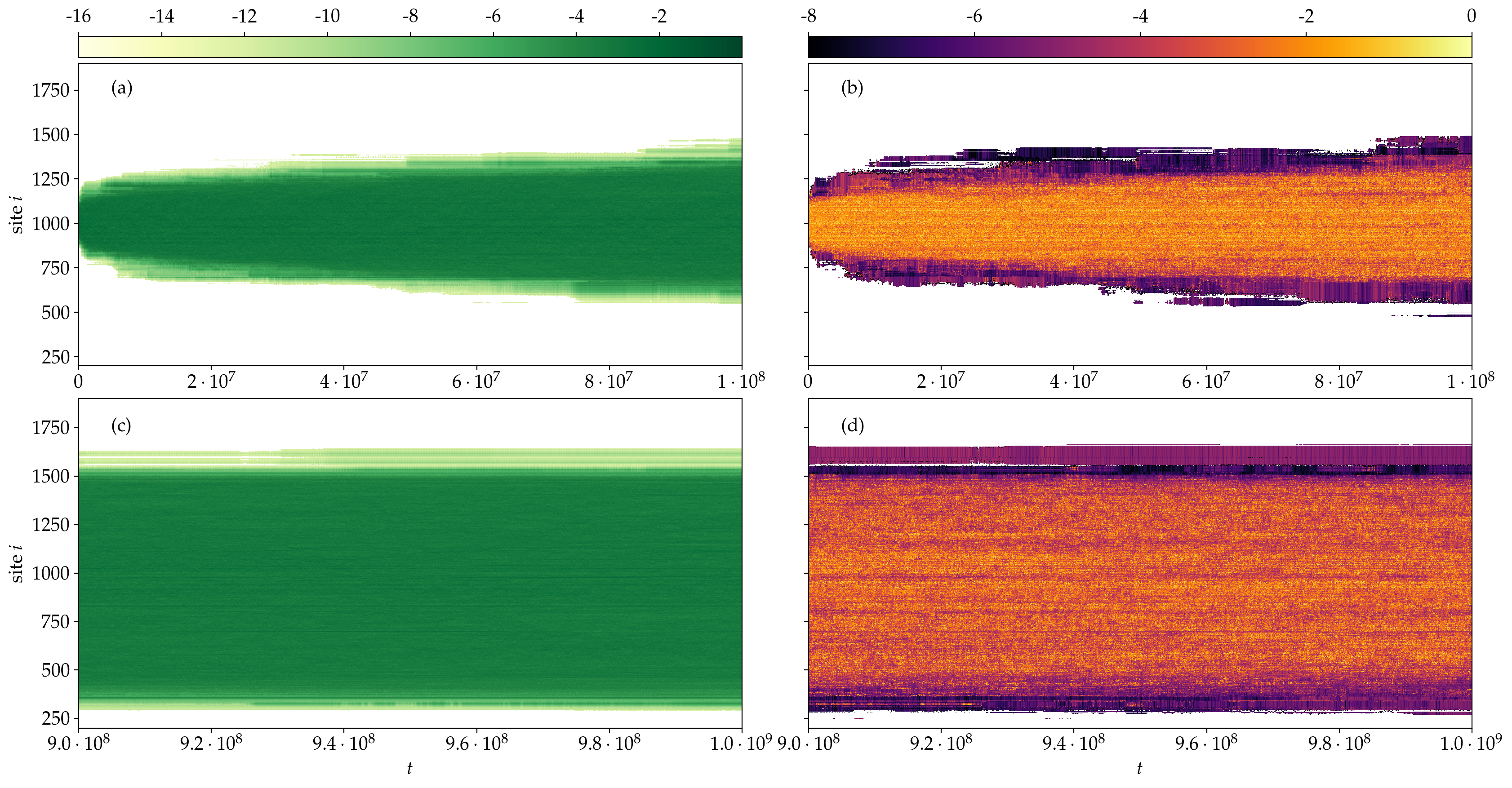}
\caption{Similar to Fig.~\ref{fig:wc1_energy_D} but for the strong chaos case SCI. }
\label{fig:sc1_energy_D}
\end{figure}

In order to investigate in depth the characteristics of chaos in the strong chaos regime, along with their differences to the weak chaos behaviors presented in Sect.~\ref{sec:num_weak}, we analyze in detail the SCI case. In our comparisons between the strong and weak chaos behaviors we will refer to the WCI case, which, as we already mentioned in Sect.~\ref{sec:num_weak}, is a good representative of the weak chaos spreading regime, since all other tested cases (like for example the WCII case of Fig.~\ref{fig:wc2}), exhibited similar behaviors. In the strong chaos regime spreading is faster than in the weak chaos case (something which is quantified by the larger exponents in the power law time evolutions of $m_2$ and $P$), so for its numerical investigation we need larger lattices in order to avoid the appearance of boundary effects. For this reason the lattice size in the SCI and SCII cases is at least doubled with respect to the weak chaos cases of  Sect.~\ref{sec:num_weak}. This faster spreading is also evident by the comparison of the energy distributions  $\xi_i$ \eqref{eq:energy_dist} in Figs.~\ref{fig:sc1_energy_D}(a) and (c) with the ones in Figs.~\ref{fig:wc1_energy_D}(a) and (c), as in the former case significantly more sites have been excited as  same times.

\begin{figure}
\centering
\includegraphics[width=\textwidth]{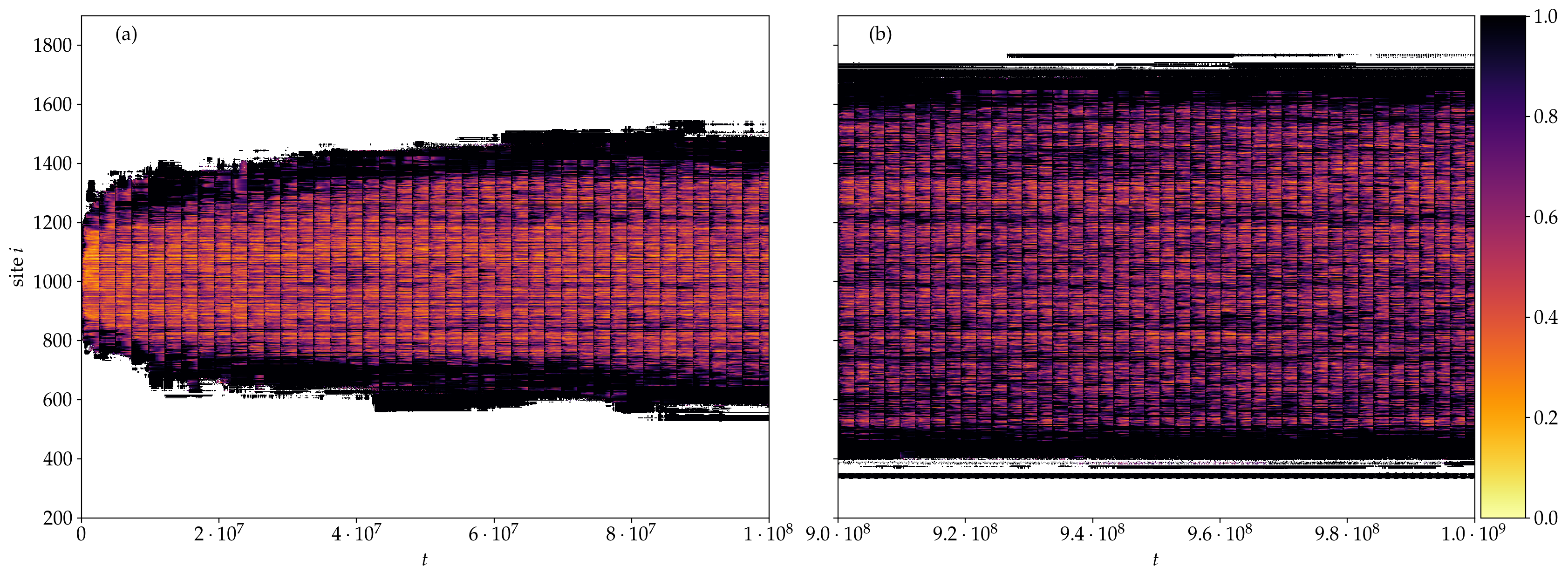}
\caption{Similar to Fig.~\ref{fig:wc1_FL} but for the strong chaos case SCI. }
\label{fig:sc1_FL}
\end{figure}

What is more interesting though, is the picture emerging by contrasting the chaotic behavior of the SCI and WCI cases through the comparison of their  $D_i$ \eqref{eq:D} distributions respectively presented in Figs.~\ref{fig:sc1_energy_D}(b), (d) and Figs.~\ref{fig:wc1_energy_D}(b), (d). In all these figures the edges of the wave packet are colored in purple and black, indicating low $D_i$ values and consequently non-chaotic behavior. On the other hand, the regions of active chaos colored in red and yellow are significantly more extended in the  SCI case, covering a larger portion of the excited part of the lattice than in the WCI case. This clear difference between the strong and weak chaos dynamical behaviors is also vividly seen by the comparison of the frequency locking $FL_i$ values presented respectively in Figs.~\ref{fig:sc1_FL} and \ref{fig:wc1_FL}. From the results of Fig.~\ref{fig:sc1_FL}(a) it is obvious that from the first stages of the evolution chaos (i.e.~regions with $FL_i \neq 1$) is quite extended, covering almost the whole wave packet apart from a rather narrow area at its edges where regular behavior (i.e.~points corresponding to  $FL_i = 1$, which are colored in black) appears. This arrangement persists also at later stages of the evolution as can be seen in Fig.~\ref{fig:sc1_FL}(b).

\begin{figure}
\centering
\includegraphics[width=\textwidth]{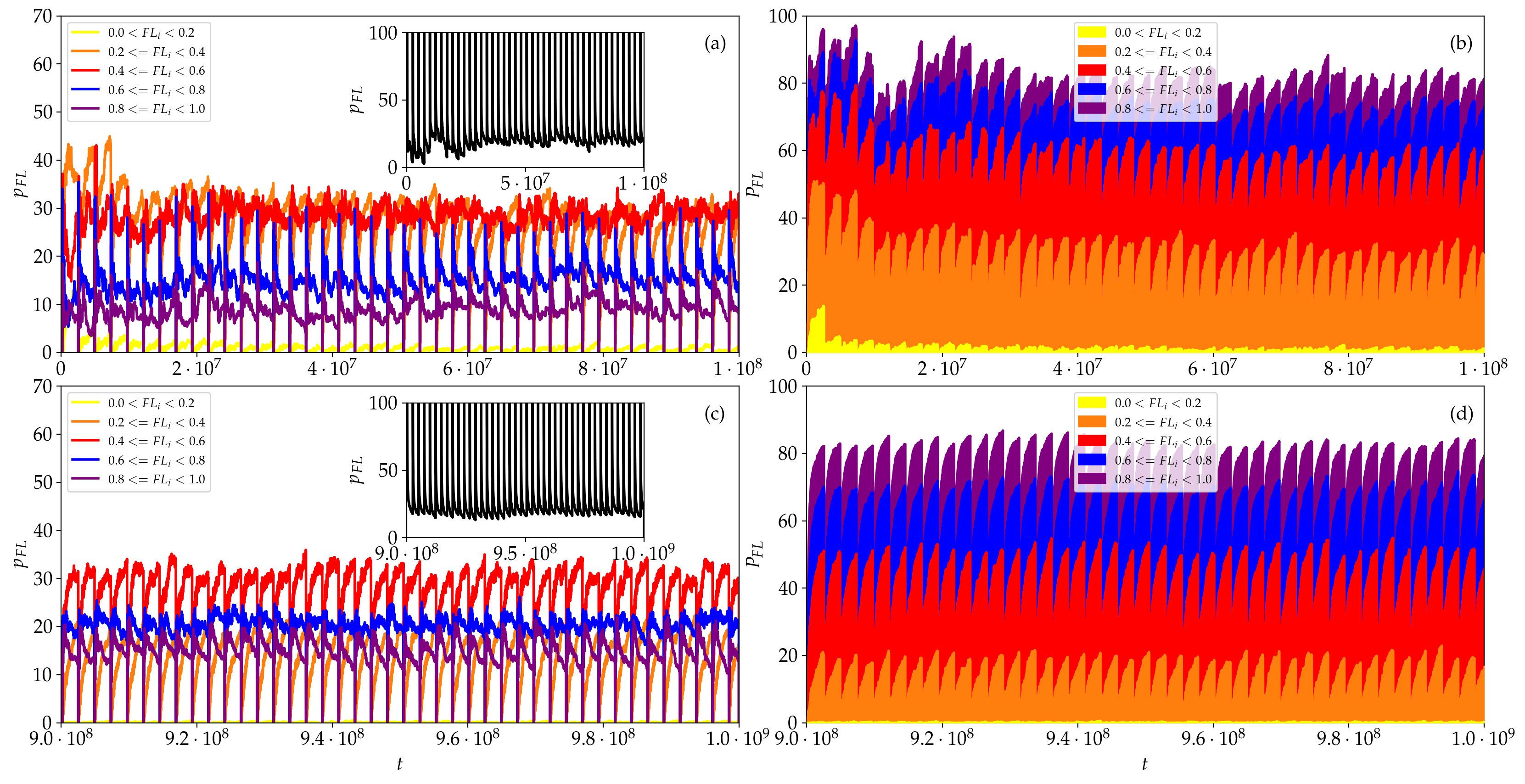}
\caption{Similar to Fig.~\ref{fig:wc1_FL_percent} but for the strong chaos case SCI. }
\label{fig:sc1_FL_percent}
\end{figure}

It is worth noting though that the area of strongly chaotic regions in Fig.~\ref{fig:sc1_FL}(b), corresponding to very low $FL_i$ values (i.e.~points colored in orange and yellow) is decreased with respect to  Fig.~\ref{fig:sc1_FL}(a). In order to understand better this behavior we present in Fig.~\ref{fig:sc1_FL_percent} results for the percentages $p_{FL}$ of sites exhibiting $FL_i$ values in particular intervals [Figs.~\ref{fig:sc1_FL_percent}(a) and (c)], along with the related accumulated percentages $P_{FL}$ [Figs.~\ref{fig:sc1_FL_percent}(a) and (c)], in a similar fashion to Fig.~\ref{fig:wc1_FL_percent}. The analysis of these results, as well as their comparison with the outputs presented in  Fig.~\ref{fig:wc1_FL_percent} for the WCI case, reveal some interesting characteristics about the chaotic behavior of the strong chaos regime, along with some important differences with respect to the weak chaos case. More specifically, the percentage of sites showing consistent regular behavior, i.e.~$FL_i=1$ after some initial fluctuations [see inset of Fig.~\ref{fig:sc1_FL_percent}(a)] more or less stabilize its value at $p_{FL} \approx 20 \%$,   something which remains true also at the later stages of the evolution, as is seen in the  inset of Fig.~\ref{fig:sc1_FL_percent}(c). This behavior is also revealed by the practical constancy of the height of the colored area in Figs.~\ref{fig:sc1_FL_percent}(b) and (d), which indicate the $P_{FL}$ values of sites with $FL_i \neq 1$. A similar behavior was also observed for the weak chaos regime, i.e.~the percentage $p_{FL}$ of sites with $FL_i=1$ saturates to an almost constant value [see insets of Figs.~\ref{fig:wc1_FL_percent}(a) and (c)], but that value is significantly larger,  $p_{FL} \approx 60 \%$, than the one obtained for the SCI case. This difference in $p_{FL}$ values actually quantifies the observation that the width of the regularly behaving zone at the edges  of the wave packet (i.e.~points colored in purple and black in Figs.~\ref{fig:sc1_energy_D}(b) and (d) and in black in Fig.~\ref{fig:sc1_FL}) is smaller than in the weak chaos case [see Figs.~\ref{fig:wc1_energy_D}(b) and (d), and Fig.~\ref{fig:wc1_FL}].

The differences between the weak and strong chaos cases go beyond the decrease of the extent of the regular component in the latter case (or equivalently the increase of the chaotic portion), as also the chaotic parts of the wave packet exhibit disparate characteristics. In particular, from Figs.~\ref{fig:sc1_FL_percent}(a) and (b) we see that at the initial stages of the evolution the percentage of strongly chaotic regions, i.e.~sites with $FL_i <0.4$, shows a tendency to slightly decrease. This tendency becomes more evident in  Fig.~\ref{fig:sc1_FL_percent}(b) where the height reached by the orange colored area moderately lessens attaining values $P_{FL} \approx 30 \%$ towards the end of the depicted time interval. These levels are significantly larger than the  $P_{FL} \approx 3 \%$ values observed in the weak chaos case WCI for the same time interval in Fig.~\ref{fig:wc1_FL_percent}(b). At later stages of the evolution [Figs.~\ref{fig:sc1_FL_percent}(c) and (d)] the fractions of sites with $FL_i$ values at different intervals tend to oscillate around some well defined levels. In contrast to the weak chaos case of Fig.~\ref{fig:wc1_FL_percent}(d) where $P_{FL} \lesssim 3 \%$ for sites with $FL_i <0.4$, in the strong chaos case we see that the percentages of these highly chaotic sites oscillates around much higher levels as  $P_{FL} \approx 20 \%$. 

The higher degree of chaos  in the strong chaos case, in comparison to the weak chaos regime, is also evident by the elevated $p_{FL}$ values of sites exhibiting moderate chaotic behavior. More specifically, we see that $p_{FL}\approx 30 \%$ and $p_{FL}\approx 20 \%$ respectively for the $0.4 \leq FL_i <0.6$ [red curve in Fig.~\ref{fig:sc1_FL_percent}(c)] and $0.6 \leq FL_i <0.8$ [blue curve in Fig.~\ref{fig:sc1_FL_percent}(c)] cases of the strong chaos regime, while we had $p_{FL}\approx 8 \%$ [red curve in Fig.~\ref{fig:wc1_FL_percent}(c)] and $p_{FL}\approx 10 \%$ [blue curve in Fig.~\ref{fig:wc1_FL_percent}(c)] for these two cases in the weak chaos regime. Furthermore, although the percentages of low strength chaos, i.e.~sites having  $0.8 \leq FL_i <1$, are similar in both the strong and weak chaos cases, with $p_{FL}\approx 17 \%$ [purple curves in respectively  Figs.~\ref{fig:sc1_FL_percent}(c) and \ref{fig:wc1_FL_percent}(c)], it is worth noting that this fraction is lower (higher) than the ones corresponding to the $0.4 \leq FL_i <0.6$ and $0.6 \leq FL_i <0.8$ cases in the strong (weak) chaos regime. This behavior indicates again the higher degree of chaos appearing in the strong chaos case. 

\begin{figure}
\centering
\includegraphics[width=\textwidth]{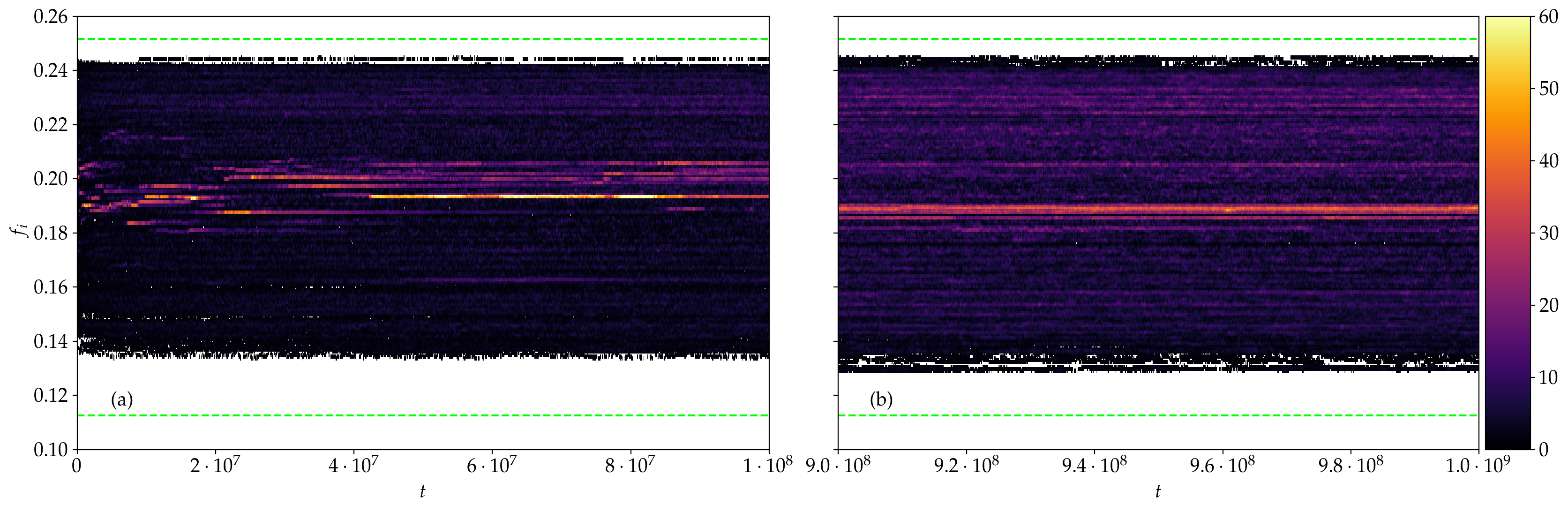}
\caption{Similar to Fig.~\ref{fig:wc1_bin} but for the strong chaos case SCI. }
\label{fig:sc1_bin}
\end{figure}

In Fig.~\ref{fig:sc1_bin} we see the time evolution of the computed fundamental frequencies $f_i$ for the SCI case. More accurately for the creation of this figure we register the frequency bins in which site frequencies reside in the time intervals $0 \leq t \leq 10^8$ [Fig.~\ref{fig:sc1_bin}(a)] and  $9 \cdot 10^8 \leq t \leq 10^9$ [Fig.~\ref{fig:sc1_bin}(b)], with each occupied bin colored according to its $FS$ value (i.e.~the number of sites having frequencies in that bin), in analogy to Fig.~\ref{fig:wc1_bin}. Similarly to what was observed in the weak chaos case in Fig.~\ref{fig:wc1_bin}, the number of excited frequencies increases in time, although this increase is not as profound as in the WCI case. The striking difference between Figs.~\ref{fig:wc1_bin} and \ref{fig:sc1_bin} is that in the strong chaos case significantly more frequencies are excited than in the weak chaos regime, for both depicted time intervals. This is in perfect agreement with the theoretical prediction that the strong chaos spreading behavior is related to a widespread excitation of normal mode frequencies \cite{SF10,LBKSF10,F10,BLSKF11}. Another important difference is that in the strong chaos case many more frequencies influence the dynamics than in the weak chaos case. This is vividly seen in  Fig.~\ref{fig:sc1_bin}(b) where a lot of frequency bins, throughout the whole frequency spectrum, are highly populated (i.e.~having large $FS$ values and consequently colored in red and yellow). Nevertheless, again a concentration of bins with high $FS$ values at the central region of the frequency band is observed.

\begin{figure}
\centering
\includegraphics[width=\textwidth]{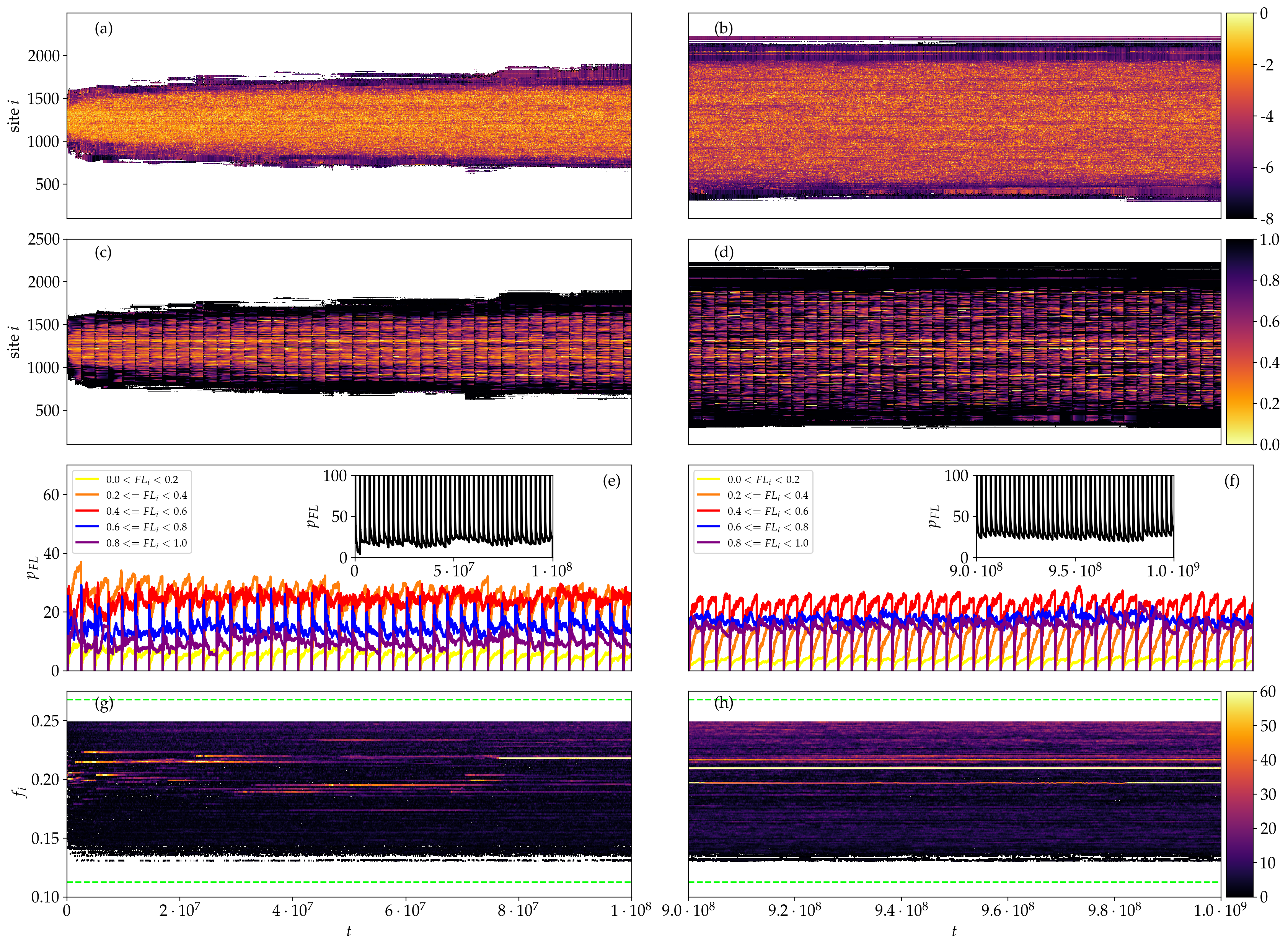}
\caption{Similar to Fig.~\ref{fig:wc2} but for the strong chaos case SCII. It is worth noting the different frequency band of the linear DKG system \eqref{eq:fn_boundaries}, which is indicated by the horizontal thick dashed green lines in panels (g) and (h), as in this case $W=3$ in contrast to $W=4$ in all other studied cases. }
\label{fig:sc2}
\end{figure}

The results presented in Figs.~\ref{fig:sc1_energy_D}--\ref{fig:sc1_bin} for the SCI case are typical for the strong chaos spreading dynamical regime, as similar behaviors have been obtained for all other tested strong chaos cases. The explicit results for one such case, the one we named SCII in Sect.~\ref{sec:num_res}, are presented in Fig.~\ref{fig:sc2}. It is worth noting that the width of the frequency band \eqref{eq:fn_boundaries} of the system's normal modes, denoted by the horizontal dashed green lines in   Figs.~\ref{fig:sc2}(g) and (h), is larger than the ones related to all other studied cases having $W=4$, as $W=3$ for the SCII case. Nevertheless, the qualitative features observed in Figs.~\ref{fig:sc2}(g) and (h) are similar to the ones seen in Fig.~\ref{fig:sc1_bin}.

\subsection{Selftrapping dynamical regime}\label{sec:num_self}

Although the main objective of our work is to understand the spatiotemporal behavior of chaos for energy propagation in disordered lattices, as well as to identify the differences between the weak and strong chaos spreading regimes, we also consider in our study, mainly for completeness' sake, a case belonging to the selftrapping regime. This is the STI case mentioned in Sect.~\ref{sec:num_res}, whose dynamics has been studied in \cite{FKS09,SKKF09}.

The existence of the selftrapping dynamical regime was theoretically predicted for the 1D DDNLS system in \cite{KKFA08} and was further discussed in \cite{SKKF09}. According to those studies, in the case of single site excitations, and for strong enough nonlinearities, the induced nonlinear frequency shift moves the frequencies of some of the excited oscillators outside the model's linear spectrum, resulting in the localization of a part of the wave packet, with the remainder spreading subdiffusively. Although the theoretical prediction of this behavior was done only in the case of the 1D DDNLS model, the selftrapping regime was numerically observed both for the 1D DDNLS and DKG systems in several publications, e.g.~\cite{FKS09,SKKF09,SF10,LBKSF10,BLSKF11}, mainly through the monitoring of wave packet profiles and the computation of $m_2$ and $P$, with the latter remaining practically constant. Thus, these numerical studies denoted the generality of the selftrapping behavior.

\begin{figure}
\centering
\includegraphics[width=\textwidth]{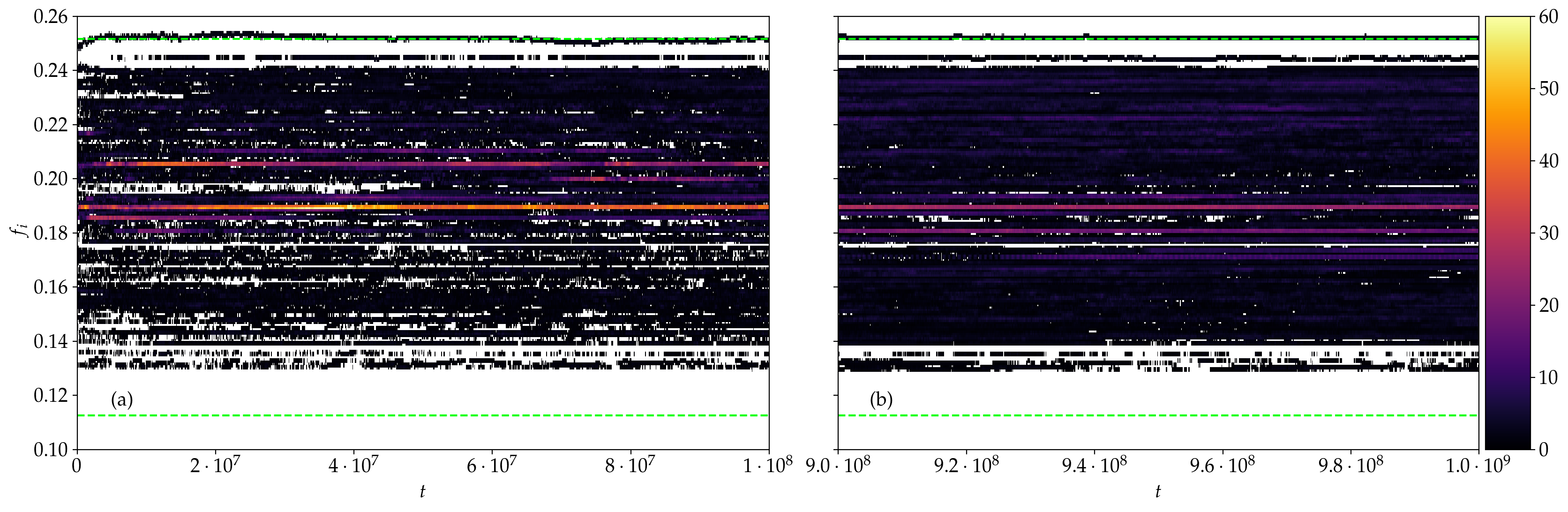}
\caption{Similar to Fig.~\ref{fig:wc1_bin} but for the selftrapping case STI.  }
\label{fig:st1}
\end{figure}

The time evolution of the STI case's fundamental frequencies, along with their $FS$, given in Fig.~\ref{fig:st1}, clearly show, to the best of our knowledge for the first time, the existence of lattice sites having frequencies outside the frequency band of the linear model (identified by the horizontal, dashed green lines in  Fig.~\ref{fig:st1}), something which is a basic ingredient for the theoretical description  of the selftrapping regime. Thus, our numerical results provide strong additional evidences of the correctness of the developed theoretical framework.

Furthermore, it is worth noting some similarities between Fig.~\ref{fig:st1} and the behaviors observed for the weak chaos case [Fig.~\ref{fig:wc1_bin} and Figs.~\ref{fig:wc2}(g) and (h)], as in both cases fewer frequencies are excited with respect to the strong chaos case  [Fig.~\ref{fig:sc1_bin} and Figs.~\ref{fig:sc2}(g) and (h)]. In addition, in both the weak chaos and selftrapping cases a rather small number of frequencies have large $FS$ values, corresponding to red and yellow colored points in Figs.~\ref{fig:wc1_bin}, \ref{fig:wc2}(g) and (h) (weak chaos) and in Fig.~\ref{fig:st1} (selftrapping). Moreover, the number of excited frequencies increase as time grows, as the size of the white region in Fig.~\ref{fig:st1}(b) for $9 \cdot 10^8 \leq t \leq 10^9$ is smaller than the one of the initial phase of the evolution in Fig.~\ref{fig:st1}(a).

\section{Summary and discussion}\label{sec:summary}

Based on results obtained by the FMA of the evolution of initially localized energy excitations we studied the features of spatiotemporal chaos in a prototypical model of 1D disordered nonlinear lattices, namely the DKG system. We focused our investigation on well-established dynamical behaviors leading to the destruction of Anderson localization, which have been investigated in the past by different approaches \cite{FKS09,SKKF09,LBKSF10,SF10,F10,BLSKF11,BLGKSF11,SGF13,ABSD14,ASBF17,SMS18}: the so-called weak and strong chaos spreading regimes. In particular, we studied in detail the characteristics and the evolution of chaos in these two cases, as well as explored the differences encountered between these regimes.

Using the relative change $D_i$ \eqref{eq:D} of the computed fundamental frequencies in two successive time windows as an indicator of chaos for each lattice site [Figs.~\ref{fig:wc1_energy_D}(b), (d); \ref{fig:wc2}(a), (b); \ref{fig:sc1_energy_D}(b), (d) and \ref{fig:sc2}(a), (b)] we showed that, in both the weak and strong chaos cases, chaotic behavior mainly appears at the central regions of the wave packet, where also the energy density $\xi_i$ \eqref{eq:energy_dist} [Figs.~\ref{fig:wc1_energy_D}(a), (c) and  \ref{fig:sc1_energy_D}(a), (c)] is relatively large, while sites at the edges of the wave packet forms a zone of regular motion. The width of this zone is larger in the weak chaos case, which consequently means that the chaotic component of the wave packet is more extended in the strong chaos regime.

In order to quantify further these findings, and to also investigate in more detail the features of the chaotic part of the dynamics, we followed the progression of a quantity we named frequency locking ($FL_i$), which is measuring the degree of practical frequency constancy (denoting non-chaotic behavior), of each oscillator in time windows of fixed length [Figs.~\ref{fig:wc1_FL}; \ref{fig:wc1_FL_percent}; \ref{fig:wc2}(c), (d), (e), (f);  \ref{fig:sc1_FL}; \ref{fig:sc1_FL_percent}; \ref{fig:sc2}(c), (d), (e), (f)]. The analysis of our $FL_i$ computations showed that the percentage of non-chaotic oscillators in the wave packet is about 3 times larger in the weak chaos regime with respect to the strong chaos case. Obviously this means that the fraction of oscillators behaving chaotically is much larger in the strong chaos regime. Apart from this important difference, our study showed that the percentage of strongly chaotic sites, i.e.~oscillators continuously exhibiting many and large changes in their frequencies and consequently having low $FL_i$ values (typically $FL_i<0.4$), is much higher in the strong chaos case, usually around 5 times higher than in the weak chaos one. These observations, legitimize and further support, in some sense, the naming of the weak and strong chaos regimes \cite{F10,SF10,LBKSF10}, as in the latter case chaotic behavior is indeed more pronounced and extended.

An additional interesting feature we noticed is that, in both spreading regimes, although the percentage of chaotic oscillators eventually remains almost fixed (because the percentage of regularly behaving sites with $FL_i=1$ practically saturates to a more or less constant value [insets of Figs.~\ref{fig:wc1_FL_percent}(a), (c); \ref{fig:wc2}(e), (f);   \ref{fig:sc1_FL_percent}(a), (c) and \ref{fig:sc2}(e), (f)]) the fraction of highly chaotic oscillators (sites having e.g.~$FL_i<0.4$) decreases in time. Thus, chaos persists although it becomes weaker in time. This observation is in agreement with the picture emerging from the computation of the ftMLE $\Lambda(t)$ [Figs.~\ref{fig:wc1_m2pL}(c) and \ref{fig:sc1_m2pL}(c)], that as time passes, the strength of chaos [quantified by the value of $\Lambda(t)$] decreases, but nevertheless without showing any tendency to crossover to regular dynamics (which corresponds to $\Lambda(t) \propto t^{-1}$), as was also shown in various studies through the computation of other quantities, like for example $m_2$ \cite{F10,SF10,LBKSF10,BLSKF11}, $q$-Gaussians \cite{ABSD14,ASBF17} and $\Lambda$ \cite{SGF13,SMS18}.

All these results allow us to better understand the mechanisms of energy spreading and the process of chaotic destruction of Anderson localization for the case of initially localized excitations in disordered lattices. The constant energy spreads in time to more sites as oscillators at the edges of the wave packet start acquiring energy and begin their motion, initially in a regular, non-chaotic fashion. As time passes these oscillators gain more energy and eventually become chaotic, and get incorporated in the significantly excited part of the wave packet, not belonging any more to the regularly-behaving edges. As additional oscillators get excited, the energy is distributed among more sites reducing in this way the amount of energy per particle in the excited part of the lattice. Thus, although the number of chaotic oscillators is continuously increasing the strength of their chaotic behavior is lessened.

A significant outcome of our study is the finding of the excited frequencies in the dynamics and, more importantly, the identification of the ones which influence the most the system's behavior [Figs.~\ref{fig:wc1_bin}; \ref{fig:wc2}(g), (h); \ref{fig:sc1_bin} and \ref{fig:sc2}(g), (h)], through the computation of the number of oscillators moving according to these frequencies (a quantity we refer to as the related frequency strength, $FS$). A remarkable difference between the weak and strong chaos regimes is that in the latter case a significantly larger number of frequencies is excited, even from the first stages of the evolution. This contrast remains valid for the whole duration of our numerical simulations, with more frequencies been activated as time passes, although a higher increase rate is observed in the weak chaos regime. The excitation of more frequencies in the strong chaos case is in accordance to the extended chaotic interactions of the system's normal modes, which were theorized to take place in this case \cite{SF10,LBKSF10,F10, BLSKF11}.

Furthermore, it is worth mentioning that in our study we also provided some significant numerical results for the selftrapping dynamical regime, where a large part of the wave packet remains localized. Our findings (Fig.~\ref{fig:st1}) confirm the shifting of frequencies outside the normal mode frequency band, a phenomenon which is at the core of the theoretical treatment of this dynamical behavior.

Apart from understanding in more depth the chaotic spreading processes in disordered nonlinear lattices, the numerical approaches implementing in our study (FMA, variations of fundamental frequencies, $FL$ and $FS$) can be used to follow possible subtle changes in the local dynamical evolution of multidimensional systems.  The overall characterization of chaos through the computation of some chaos indicators, like the MLE, depends on the collective behavior of the system and fails to identify localized differences in the dynamics, like for example the appearance of localized chaotic hot spots \cite{TSL14}. The study of the properties of the deviation vector distribution (DVD) related to the computation of the MLE has already been implemented as a mean to tackle this problem \cite{SGF13,SMS18,NTRSA19,MFAERF19}. The numerical methods presented here can be implemented as alternative and complementary approaches for the same purpose.

\nonumsection{Acknowledgments}
Ch.~Skokos acknowledges support by  the University of Cape Town Research Committee (URC) and thanks the Max Planck Institute for the Physics of Complex Systems in Dresden, Germany, for its hospitality during his visits there in 2018 and 2019, when parts of this work were carried out. E.~Gerlach thanks the Centre for Information Services and High Performance Computing (ZIH) at the Technische Universität (TU) Dresden, where part of the computation for this work was done.  S.~Flach acknowledges support by the Institute for Basic Science, Project Code (IBS-R024-D1), and thanks the Max Planck Institute for the Physics of Complex Systems (Dresden) and the New Zealand Institute for Advanced Studies (Auckland) for hospitality during visits
when parts of this work were carried out. We also thank J.~Laskar for making available his NAFF code.




\begin{thebibliography}{9}

\bibitem[Anderson(1958)]{A58} Anderson, P. W. [1958] ``Absence of diffusion in certain random lattices'',  {\it Phys. Rev.} {\bf 109}, 1492--1505.

\bibitem[Antonopoulos {\it et al.}(2014)]{ABSD14} Antonopoulos, Ch., Bountis, T., Skokos, Ch. \& Drossos, L. [2014] ``Complex statistics and diffusion in nonlinear disordered particle chains'',  {\it Chaos} {\bf 24}, 024405.

\bibitem[Antonopoulos {\it et al.}(2017)]{ASBF17} Antonopoulos, Ch., Skokos, Ch., Bountis, T. \& Flach, S. [2017] ``Analyzing chaos in higher order disordered quartic-sextic Klein-Gordon lattices using $q$-statistics'',  {\it Chaos Sol. Fract.} {\bf 104}, 129--134.

\bibitem[Aubry(2011)]{A11} Aubry, S. [2011] ``KAM tori and absence of diffusion of a wave-packet in the 1D random DNLS model'', {\it Int. J. Bifurcation  Chaos} {\bf 21}, 2125--2145.

\bibitem[Benettin {\it et al.}(1980a)]{BGGS80a} Benettin, G., Galgani, L., Giorgilli, A. \& Strelcyn, J.-M. [1980a] ``Lyapunov characteristic exponents for smooth dynamical systems and for Hamiltonian systems; a method for computing all of them. Part 1: Theory'',  {\it Meccanica} {\bf 15}, 9--20.	

\bibitem[Benettin {\it et al.}(1980b)]{BGGS80b} Benettin, G., Galgani, L., Giorgilli, A. \& Strelcyn, J.-M. [1980b] ``Lyapunov characteristic exponents for smooth dynamical systems and for Hamiltonian systems; a method for computing all of them.  Part 2: Numerical application'',  {\it Meccanica} {\bf 15}, 21--30.

\bibitem[Blanes {\it et al.}(2013)]{BCFLMM13}	Blanes, S., Casas, F., Farr{\'e}s, A.,  Laskar, J., Makazaga, J.  \& Murua, A. [2013] ``New families of symplectic splitting methods for numerical integration in dynamical astronomy'', {\it Appl. Numer. Math.} {\bf 68}, 58--72.	

\bibitem[Bodyfelt {\it et al.}(2011a)]{BLSKF11} Bodyfelt, J. D., Laptyeva, T. V., Skokos, Ch., Krimer, D. O. \& Flach, S. [2011a] ``Nonlinear waves in disordered chains: probing the limits of chaos and spreading'', {\it Phys. Rev. E} {\bf 84}, 016205.

\bibitem[Bodyfelt {\it et al.}(2011b)]{BLGKSF11} Bodyfelt, J. D., Laptyeva, T. V., Gligoric, G., Krimer, D. O., Skokos, Ch. \& Flach, S. [2011b] ``Wave interactions in localizing media - a coin with many faces'', {\it Int. J. Bifurcation  Chaos} {\bf 21}, 2107--2124.	

\bibitem[Danieli {\it et al.}(2019)]{DMMS19} Danieli, C., Many Manda, B., Mithun, T. \& Skokos, Ch. [2019] ``Computational effciency of numerical integration methods for the tangent dynamics of many-body Hamiltonian systems in one and two spatial dimensions'',  {\it Math. Eng.} {\bf 1}, 447--488.

\bibitem[Flach(2010)]{F10} Flach, S. [2010] ``Spreading of waves in nonlinear disordered media'', {\it Chem. Phys.} {\bf 375}, 548--556.

\bibitem[Flach {\it et al.}(2009)]{FKS09} Flach, S., Krimer, D. O. \& Skokos, Ch. [2009] ``Universal spreading of wave packets in disordered nonlinear systems'', {\it Phys. Rev. Lett.} {\bf 102}, 024101 (Erratum: {\it Phys. Rev. Lett.} {\bf 102}, 209903).

\bibitem[Garc\'{i}a-Mata \& Shepelyansky(2009)]{GS09}  Garc\'{i}a-Mata, I. \& Shepelyansky D. L. [2009] ``Delocalization induced by nonlinearity in systems with disorder'', {\it Phys. Rev. E} {\bf 79}, 026205.

\bibitem[Gerlach \& Skokos(2011)]{GS11} Gerlach, E. \& Skokos, Ch. [2011] ``Comparing the efficiency of numerical techniques for the integration of variational equations: Dynamical systems, differential equations and applications'', {\it Discrete \& Continuous Dynamical Systems}-Supp. 2011 (dedicated to the 8th AIMS Conference), eds. Feng, W., Feng, Z., Grasselli, M., Ibragimov, A., Lu, X., Siegmund, S. \& Voirt, J.AIMS, pp.~475--484.

\bibitem[Gerlach {\it et al.}(2012)]{GES12} Gerlach, E., Eggl, S. \&  Skokos, Ch. [2012] ``Efficient integration of the variational equations of multidimensional Hamiltonian systems: Application to the Fermi--Pasta--Ulam lattice'', {\it Int. J. Bifurcation  Chaos} {\bf 22}, 1250216.

\bibitem[Johansson {\it et al.}(2010)]{JKA10} Johansson, M., Kopidakis, G. \& Aubry, S. [2010] ``KAM tori in 1D random discrete nonlinear Schr\"{o}dinger model?'',  {\it Europhys. Lett.} {\bf 91}, 50001.

\bibitem[Kati {\it et al.}(2020)]{KYF20} Kati, Y, Yu, X. \& Flach, S. [2019] ``Density resolved wave packet spreading in disordered Gross-Pitaevskii lattices'', {\it SciPost Phys. Core} {\bf 3}, 006.

\bibitem[Kopidakis {\it et al.}(2008)]{KKFA08}  Kopidakis, G., Komineas, S., Flach, S.  \& Aubry, S. [2008] ``Absence of wave packet diffusion in disordered nonlinear systems'', {\it Phys. Rev. Lett.} {\bf 100}, 084103.

\bibitem[Kramer \& MacKinnon(1993)]{KM93} Kramer, B. \& MacKinnon A. [1993] ``Localization: theory and experiment'',  {\it Rep. Prog. Phys.} {\bf 56}, 1469--1564.

\bibitem[Laskar(1990)]{Laskar1990} Laskar, J. [1990] ``The chaotic motion of the Solar System. A numerical estimate of the size of the chaotic zones'',  {\it Icarus} {\bf 88},  pg. 266–291.

\bibitem[Laskar {\it et al.}(1992)]{LFC1992} Laskar, J., Froeschl\'{e}, and Celletti, A.  [1992] ``The measure of chaos by the numerical analysis of the fundamental frequencies. Application to the standard mapping'',  {\it Physica D} {\bf 56},  pg. 253–269.

\bibitem[Laskar(1993)]{Laskar1993} Laskar, J. [1993] ``Frequency analysis for multi-dimensional systems: global dynamics and diffusion'',  {\it Physica D} {\bf 67},  pg. 257–281.

\bibitem[Laptyeva {\it et al.}(2010)]{LBKSF10} Laptyeva, T. V., Bodyfelt, J. D., Krimer, D. O., Skokos, Ch. \& Flach, S. [2010] ``The crossover from strong to weak chaos for nonlinear waves in disordered systems'',  {\it Europhys. Lett.} {\bf 91} 30001.

\bibitem[Laptyeva {\it et al.}(2012)]{LBF12} Laptyeva, T. V., Bodyfelt, J. D. \& Flach, S. [2012] ``Subdiffusion of nonlinear waves in two-dimensional disordered lattices'',  {\it Europhys. Lett.} {\bf 98}, 60002.

\bibitem[Laptyeva {\it et al.}(2014)]{LIF14} Laptyeva, T. V., Ivanchenko, M. V. \& Flach, S. [2014] ``Nonlinear lattice waves in heterogeneous media'',  {\it J. Phys. A: Math. Theor.} {\bf 47}, 493001.

\bibitem[Many Manda {\it et al.}(2020)]{MSS20} Many Manda, B., Senyange, B. \& Skokos, Ch. [2020] ``Chaotic wave-packet spreading in two-dimensional disordered nonlinear lattices'',  {\it Phys. Rev. E} {\bf 101}, 032206.

\bibitem[Miranda Filho {\it et al.}(2019)]{MFAERF19} Miranda Filho, L. H., Amato, M. A., Elskens, Y. \&  Rocha Filho, T. M. [2019] ``Contribution of individual degrees of freedom to Lyapunov vectors in many-body systems'', {\it Comm.Nonlinear Sci. Num. Simul.} {\bf 74}, 236--247.

\bibitem[Molina(1998)]{M98} Molina, M. I. [1998] ``Transport of localized and extended excitations in a nonlinear Anderson model'',  {\it Phys. Rev. B} {\bf 58}, 12547--12550.

\bibitem[Ngapasare {\it et al.}(2019)]{NTRSA19} Ngapasare, A., Theocharis, G., Richoux, O., Skokos, Ch. \& Achilleos, V. [2019] ``Chaos and Anderson localization in disordered classical chains: Hertzian versus Fermi-Pasta-Ulam-Tsingou models'', {\it Phys. Rev. E} {\bf 99}, 032211.

\bibitem[Papaphilippou(2014)]{Y2014} Papaphilippou, Y. [2014] ``Detecting chaos in particle accelerators through the frequency map analysis method'', {\it Chaos} {\bf 24}, 024412.

\bibitem[Pikovsky \& Shepelyansky(2008)]{PS08} Pikovsky, A. S. \& Shepelyansky D. L. [2008] ``Destruction of Anderson localization by a weak nonlinearity'', {\it Phys. Rev. Lett.} {\bf 100}, 094101.

\bibitem[Pikovsky \& Polity(2016)]{PP16} Pikovsky, A. \& Politi, A. [2016] {\it Lyapunov Exponents: A Tool to Explore Complex Dynamics} (Cambridge University Press, UK).

\bibitem[Robutel \& Laskar(2000)]{RL2000} Robutel, P. \& Laskar, J. [2000] Frequency Map and Global Dynamics in the Solar System I'', {\it Icarus} {\bf 152}, 4--28.

\bibitem[Senyange \& Skokos(2018)]{SS18} Senyange, B. \& Skokos, Ch. [2018] ``Computational efficiency of symplectic integration schemes: application to multidimensional disordered Klein--Gordon lattices'',  {\it Eur. Phys. J. Spec. Top.} {\bf 227}, 625--643.	

\bibitem[Senyange \& Skokos(2021)]{SS21} Senyange, B. \& Skokos, Ch. [2021] ``Identifying localized and spreading chaos in nonlinear disordered lattices by the Generalized Alignment Index (GALI) method'',  {\it ArXiv:} 2112.02254.	

\bibitem[Senyange {\it et al.}(2018)]{SMS18} Senyange, B., Many Manda, B. \& Skokos, Ch. [2018] ``Characteristics of chaos evolution in one-dimensional disordered nonlinear lattices'', {\it Phys. Rev. E} {\bf 98}, 052229 (Erratum: {\it Phys. Rev. E} {\bf 99}, 069903).

\bibitem[Senyange {\it et al.}(2020)]{SPMS20} Senyange, B., du Plessis J.-J., Many Manda, B. \& Skokos, Ch. [2020] ``Properties of normal modes in a modified disordered Klein-Gordon lattice: From disorder to order '', {\it Nonlin. Phenom. Complex Syst.} {\bf 23}, 165--171.

\bibitem[Skokos(2010)]{S10} Skokos, Ch. [2010] ``The Lyapunov characteristic exponents and their computation'', {\it Lect. Notes Phys.} {\bf 790}, 63--135.

\bibitem[Skokos \& Flach(2010)]{SF10} Skokos, Ch. \&  Flach, S. [2010] ``Spreading of wave packets in disordered systems with tunable nonlinearity'', {\it Phys. Rev. E} {\bf 82}, 016208.

\bibitem[Skokos \& Gerlach(2010)]{SG10} Skokos, Ch. \& Gerlach, E. [2010] ``Numerical integration of variational equations'', {\it Phys. Rev. E} {\bf 82}, 036704.

\bibitem[Skokos {\it et al.}(2009)]{SKKF09} Skokos, Ch., Krimer, D. O., Komineas, S. \&  Flach, S. [2009] ``Delocalization of wave packets in disordered nonlinear chains'', {\it Phys. Rev. E} {\bf 79}, 056211 (Erratum: {\it Phys. Rev. E} {\bf 89}, 029907).

\bibitem[Skokos {\it et al.}(2013)]{SGF13} Skokos, Ch., Gkolias, I. \& Flach, S. [2013] ``Nonequilibrium chaos of disordered nonlinear waves'', {\it Phys. Rev. Lett.} {\bf 111}, 064101.

\bibitem[Tieleman {\it et al.}(2014)]{TSL14} Tieleman, O., Skokos, Ch. \& Lazarides, A. [2014] ``Chaoticity without thermalisation in disordered lattices'',  {\it Europhys. Lett.} {\bf 105}, 20001.

\bibitem[Vakulchyk {\it et al.}(2019)]{VFF19} Vakulchyk, I., Fistul, M. V. \& Flach, S. [2019] ``Wave packet spreading with disordered nonlinear discrete-time quantum walks'', {\it Phys. Rev. Lett.} {\bf 122}, 040501.



\end{thebibliography}
\end{document}